\documentclass{aa}  
\usepackage{graphicx}
\usepackage{txfonts}

\usepackage{tikz}
\usepackage{xcolor}
\usepackage{upgreek}

\newcommand{\vdisp}[0]{\ensuremath{v_\mathrm{1D}}}

\begin{document}

\title{Survival of the most compact: the life and death of satellite halos in self-interacting dark matter}
\titlerunning{Satellite halos in SIDM}

   \author{David Klemmer\inst{\ref{inst:itp}},
          Moritz S.~Fischer\inst{\ref{inst:dipc},\ref{inst:usm},\ref{inst:origins}},
          Kimberly K.~Boddy\inst{\ref{inst:wi}},
          Manoj Kaplinghat\inst{\ref{inst:uci}},
          Laura Sagunski\inst{\ref{inst:itp}}
          }
    \authorrunning{D.\ Klemmer et al.}

    \institute{
        Institute for Theoretical Physics, Goethe University, 60438 Frankfurt am Main, Germany\label{inst:itp}\\
        \email{dklemmer@itp.uni-frankfurt.de}
        \and
        Donostia International Physics Center (DIPC), Paseo Manuel de Lardizabal 4, 20018 Donostia-San Sebastian, Spain\label{inst:dipc}
        \and
        Universitäts-Sternwarte, Fakultät für Physik, Ludwig-Maximilians-Universität München, Scheinerstr.\ 1, D-81679 München, Germany\label{inst:usm}
        \and
        Excellence Cluster ORIGINS, Boltzmannstrasse 2, D-85748 Garching, Germany\label{inst:origins}
        \and
        Texas Center for Cosmology and Astroparticle Physics, Weinberg Institute, Department of Physics, The University of Texas at Austin, Austin, TX 78712, USA\label{inst:wi}
        \and
        Center for Cosmology, Department of Physics and Astronomy, University of California - Irvine, Irvine, CA 92697, USA\label{inst:uci}
    }

    \date{Received XX Month, 2026 / Accepted XX Month, 20XX}
    
    \abstract{Self-interacting dark matter (SIDM) models feature short-range interactions between dark matter (DM) particles that lead to larger diversity in the inner parts of galactic rotation curves and potentially unique gravitational lensing signatures. Satellite galaxies and dark subhalos provide a valuable testing ground for such models.
    }    
    {We develop a simulation framework to explore subhalo evolution and its gravothermal collapse for velocity- and angle-dependent self-interacting cross section in these SIDM models. Our results are essential for testing these models.}
    {We perform high-resolution $N$-body simulations, treating the host halo analytically and modelling the scattering-induced subhalo-halo interaction process realistically using virtual host particles, a central innovation of our work. We use the Eddington inversion method to accurately model the local velocity distribution in the halo. 
    Our approach is significantly less computationally expensive than simulations with a fully resolved host and subhalo, while incorporating tidal stripping and tidal heating. We test both isotropic and forward-dominated self-scattering, which represent limiting cases for the angular dependence of the self-interaction cross section.}
    {Environmental effects, especially the scattering-induced subhalo-halo interaction, have a strong impact on the subhalo evolution and drive a complex structural evolution. As a result, SIDM subhalos have a larger range of central densities and density profile slopes compared to collisionless DM.
    }
    {Our high-resolution, cost-efficient simulation framework enables modelling of SIDM subhalos in realistic environments.
    Our results highlight the necessity of accurately modelling the scattering-induced subhalo-halo interaction to predict SIDM subhalo density profiles. 
    For the range of SIDM models we investigate, the enhanced diversity in the mass profiles of subhalos would leave an observable imprint on strong lensing systems and satellite galaxies.}
  
    \keywords{methods: numerical -- dark matter -- Galaxies: dwarf}

    \maketitle

\section{Introduction}
\label{sec:intro}

Dark matter (DM) is a fundamental component of the Universe, shaping the formation and evolution of cosmic structures. Despite its well-established gravitational influence, its particle nature remains unknown. The standard collisionless cold dark matter (CDM) model, which assumes that DM interacts only through gravity, successfully explains the large-scale structure of the Universe~\citep[e.g.][]{BOSS:2016wmc,Planck:2018vyg}. Both the cold and collisionless assumptions can be tested by probing structure formation on galactic and sub-galactic scales. 

Direct, indirect and collider searches for DM particles have revealed no signals, implying that DM interacts extremely weakly with the Standard Model and perhaps even only gravitationally. There has been a renewed focus on particles in a dark sector that comprise most of the observed DM. Within this paradigm, interactions between DM particles are naturally expected, and it could be similar to the interactions we see in the Standard Model (e.g., Coulomb forces, atomic forces, nuclear forces). These interactions can reveal themselves by modifying the formation and evolution of galaxies and their DM halos. 

There are several open problems on galactic and sub-galactic scales that could be hints for the collisional nature of DM. A widely discussed problem is the observed diversity in rotation curves of field \citep{KuziodeNaray:2009oon,Oman:2015xda} and satellite galaxies \citep{Kaplinghat:2019svz, Zavala:2019sjk}. 
It is possible to explain some of this diversity with feedback from supernovae in the context of CDM simulations \citep[e.g.][]{Sales:2022ich,cruz2025dwarfdiversitylambdacdmbaryons}. 
Recent surveys---such as the Satellites Around Galactic Analogs (SAGA) survey~\citep{Mao:2024pic}, the Dark Energy Survey (DES) \citep{Bechtol:2015cbp, Drlica-Wagner:2015ufc}, and the Sloan Digital Sky Survey (SDSS) \citep{SDSS-IV:2014foz}---have advanced our understanding of satellites, providing insights into the substructure of galactic halos. Additionally, gravitational lensing, distorted by DM subhalos, offers a unique way to study the properties of satellite galaxies. Analyses of the strong-lenses SLACS0946+1006 (``Jackpot'' system) and JVAS B1938+666 suggest that the observed distortions are consistent with perturbing subhalos with central densities that are steeply rising to small radii~\citep{Minor_2021, Despali_2025, Enzi_2025, 2025ApJ...981....2M, Cao:2025eff, Li_2025, Powell_2025, Vegetti_2026}. Achieving such densities under the CDM framework is challenging, as CDM subhalos with sufficiently high concentrations are rare. An exciting possibility is that the enhanced diversity in the properties of the central dark matter densities is a signature of short-range interactions between DM particles. 

A predictive and simple model with a rich phenomenology is self-interacting dark matter (SIDM) with a large cross section for elastic scattering \citep{Spergel:1999mh, Dave:2000ar}. It is now well-established that the velocity dependence of the elastic scattering cross section is an essential feature of these models \citep[e.g.][]{Elbert:2014bma,
Kaplinghat:2015aga,
Kahlhoefer:2019oyt, 
Kaplinghat:2019svz, 
Sagunski:2020spe,
Correa_2021,
Andrade:2020lqq,
Slone_2023,
Yang:2023stn}, which makes the DM interactions qualitatively similar to those in the Standard Model. Dark sector particle physics models with a velocity-dependent self-interaction cross section for DM have been widely explored~\citep[e.g.][]{Ackerman:2008gi,Feng:2009mn, Feng:2009hw, Loeb:2010gj, Buckley:2009in, Cline:2013zca, Boddy:2014yra, Colquhoun:2020adl}. Within this class of models, self interactions between DM particles modify the evolution of DM halos by permitting heat transfer between different regions of the halo. As a result, the cuspy CDM inner density profiles of the halo are softened into isothermal cores for SIDM halos. Moreover, after generating a core during this core expansion phase, the gravothermal evolution of SIDM halos eventually yields highly concentrated inner cores during the core-collapse phase \citep{Balberg:2002ue, Koda:2011yb}. 

It was pointed out that this core collapse phase can proceed more efficiently in subhalos~\citep{Nishikawa_2020}, which sparked a renewed interest in this possibility due to the fact that it can increase the range of central DM densities in satellite galaxies~\citep{Kahlhoefer:2019oyt}. This core evolution enables SIDM to naturally explain the large variations seen in galactic rotation curves \citep[e.g.][]{Ren:2018jpt, Nadler:2023nrd, Yang_2023b, Ragagnin:2024deh, Fischer_2026a}. DM subhalos with very dense cores also offer an explanation for the recent lensing anomalies \citep{Minor_2021, Enzi_2025, Li_2025}. For a detailed overview of SIDM models and motivations, see the review articles by \citet{Tulin:2017ara, Adhikari_2022}.

The ability to reliably compare SIDM predictions with observations depends on modelling these systems appropriately. Satellite galaxies are particularly difficult to model in state-of-the-art cosmological or idealized $N$-body simulations. While $N$-body simulations can accurately model large-scale structure, capturing the detailed dynamics of smaller satellite galaxies within the same simulation is challenging due to the wide range of scales involved~\citep[e.g.][]{Barry:2023ksd}. The primary issue is the extreme mass difference between the satellite and the host galaxy: to accurately capture the SIDM dynamics of both systems within a single simulation, it is necessary to use numerical particles of the same mass for both the host and the satellite.
Resolving the much less massive satellite at high fidelity demands a very large number of particles for the host. As a result, simulating both the host galaxy and its satellites at high resolution requires enormous computational resources, significantly limiting the number of systems and DM models that can be studied.

In this paper, we investigate how SIDM affects the DM structures of dwarf satellite galaxies. We employ an efficient simulation method that focuses only on the satellite galaxies, treating the gravitational potential of the host galaxy analytically \citep[for example, similar to][]{Kahlhoefer:2019oyt, Sameie_2020}. This approach reduces computational costs and enables us to resolve the low-mass satellites.
Additionally, we implement an advanced scattering routine that models non-gravitational interactions between DM particles in the host and satellite halo~\citep{Kummer:2017bhr}. In contrast to the method by \citet{Zeng_2022}, our numerical approach uses virtual host particles with a velocity distribution obtained via Eddington inversion. Moreover, it supports multiple scattering per time step, and can handle any velocity- and angular-dependent cross section. 

In recent work, the mass-loss due directly to scattering between particles in the host and satellite halos has been variously labelled as ``evaporation''~\citep{Kummer:2017bhr,Zeng_2022} or ``ram-pressure evaporation''~\citep{Slone_2023}. The use of the terms ram-pressure and evaporation is not ideal. As discussed in these papers, the interactions between the particles in the subhalo and the halo happen in the long mean-free-path regime for practical scenarios, and subhalos do not experience ram-pressure stripping in the hydrodynamic sense. Instead, individual scatterings can transfer sufficient energy to unbind subhalo particles, leading to a gradual {\em scattering-induced} mass loss at a rate set by the local momentum flux and the momentum-transfer cross section at relative velocities around the orbital velocity~\citep{Kummer:2017bhr}. Therefore, ram-pressure stripping formulae used for gases in typical astrophysical settings may not be used for this phenomenon. 

The use of the term ``evaporation'' is also not technically accurate because the mass loss discussed above is neither a surface phenomenon nor a cooling process. For mass loss due to large momentum exchanges with host halo particles, the probability of expelling a particle from the subhalo is practically the same everywhere in the subhalo if orbital velocities are much larger than the subhalo velocity dispersion. 

In addition to the mass loss, this process also impacts the subhalo in other ways. The particles in the subhalo that have scattered with the main halo particles but have not escaped would lead to heating of the subhalo as well as a drag force~\citep{Kummer:2017bhr}. For clarity, we refer to this process in SIDM models as the \textbf{scattering-induced subhalo–halo interaction (SSHI).} 

We investigate the evolution of both CDM and SIDM subhalos in three different orbital configurations: one with an eccentricity of $2/3$ and two with eccentricities of $1/3$ but different pericentre distances.
We consider two different velocity-dependent SIDM cross sections, which we take to be either isotropic or far-forward dominated.
Running the evolution of subhalos under both CDM and SIDM allows us to highlight how SIDM influences the evolution and structural properties of the subhalos. By simulating both isotropic and forward-dominated self-scattering, we bracket the possibilities for the angular dependence of the self-interaction cross section.

We focus on four key quantities: the projected density slope and projected mass of the subhalos, the maximum circular velocity, and the corresponding radius at which this velocity occurs. We demonstrate that SIDM can produce a significantly broader diversity in those observables of DM subhalos compared to CDM.
This diversity arises not only from the internal SIDM-driven evolution, but also from external influences on the subhalo. Our key result is that accurately capturing these environmental effects is crucial for modelling the evolution of SIDM subhalos. In particular, the SSHI process can strongly influence their evolution. Those internal and environmental effects can produce an SIDM subhalo with a steep central density that can contribute to a wider diversity of galactic rotation curves among subhalos and provide an explanation for the disturbed strong lenses.

The remainder of this paper is structured as follows. In Sect.~\ref{sec:SubhaloEvoOverview}, we discuss the mechanisms shaping the properties of satellite galaxies in SIDM. In Sect.~\ref{sec:numerical_method}, we describe our numerical treatment for our simulations, including the implementation of the SSHI process model.
In Sect.~\ref{sec:evolution}, we analyse the results of our simulations, which include the evolution of both isolated halos and satellite halos that interact with the environment of their host halo.
In Sect.~\ref{sec:discussion}, we discuss the limitations of our work and its implications.
Finally, we conclude in Sect.~\ref{sec:conclusion}.
Additional information is provided in the Appendices.

\section{Subhalo evolution overview}
\label{sec:SubhaloEvoOverview}
\begin{figure*}
    \centering
    \includegraphics[width=\columnwidth]{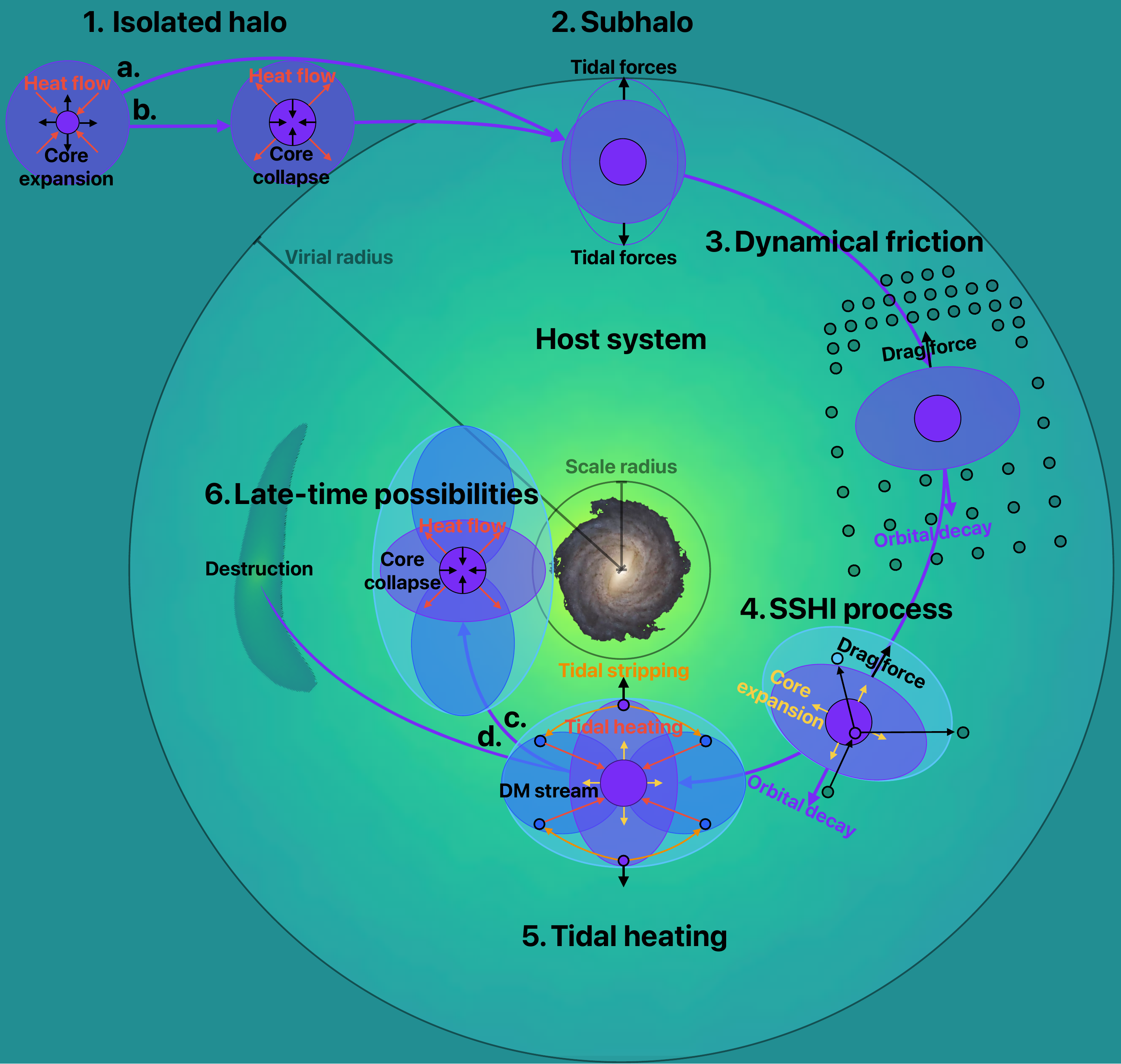}
    \includegraphics[width=\columnwidth]{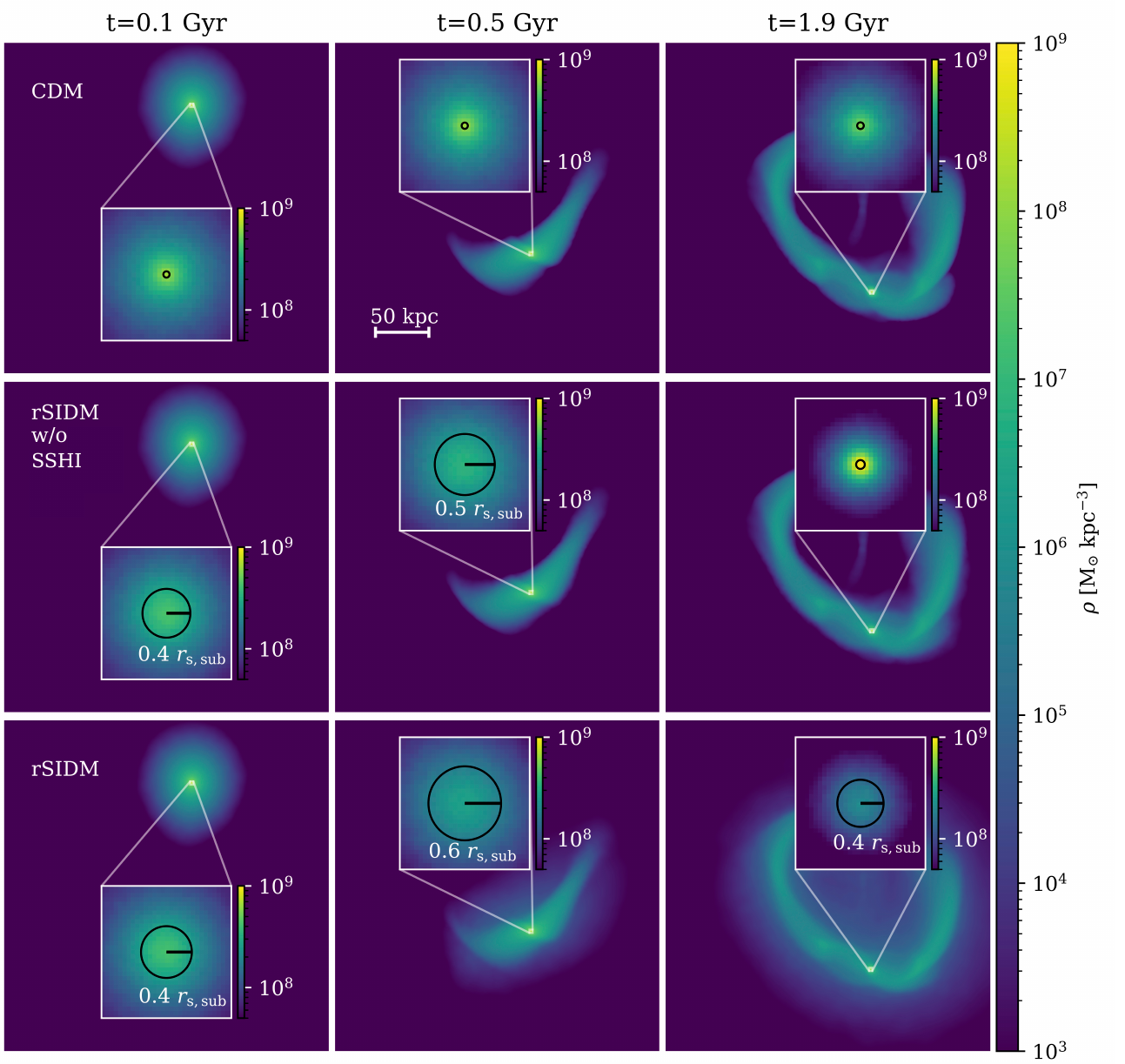}
    \caption{Left panel: Illustration of the evolution and the mechanisms acting on an isolated SIDM halo, which becomes bound to a host system over time.
    Right panel: 
    Density map in the orbital plane of a subhalo on an elliptical orbit between $r_\mathrm{apocenter}=2 \, r_\mathrm{s, \, host}$ and $r_\mathrm{pericenter}= r_\mathrm{s, \, host}$. The radii $r_\mathrm{s, \, host}$ and $r_\mathrm{s, \ sub}$ are the scale radius of the initial NFW host halo and subhalo. The top row shows the evolution of a CDM subhalo, the middle row shows an SIDM subhalo without the SSHI process, and the bottom row shows the evolution with the SSHI process. Our simulations of both SIDM subhalos use an isotropic velocity-dependent cross section.}
    \label{fig:SubhaloEvolution}
\end{figure*}
Understanding the physical processes that govern the evolution of satellite galaxies in the context of SIDM is essential for making accurate predictions about their structural and dynamical properties. This section outlines the key evolutionary stages/processes of a DM subhalo in an SIDM framework, beginning with its isolated development and continuing through its transformation into a satellite within a host halo.
We assume the isolated subhalo has an initial Navarro-Frank-White (NFW) density profile \citep{Navarro:1995iw}.
The evolutionary stages are depicted in the left panel of Fig.~\ref{fig:SubhaloEvolution} and described below.
\begin{enumerate}
\item Initially, the DM halo is isolated and undergoes core expansion due to self-interactions. The self-interaction induces a heat flow inward, reducing the central density and increasing the central velocity dispersion, gradually forming a core with a constant density. The halo eventually reaches its maximum core size, and the heat flow reverses direction. The outward heat flow causes the core size and mass to shrink, but the density and velocity dispersion grow, resulting in the runaway process of core collapse.
The isolated halo can cross the virial radius of a host halo while (a) still in the core expansion phase or (b) in the core collapse phase.

\item Once the isolated halo is in the vicinity of the host system, it enters the host’s gravitational environment and is elongated and partially stripped by tidal forces.

\item Simultaneously, the gravitational interaction with the local density of the host leads to dynamical friction \citep{Chandrasekhar_1943}, which causes the subhalo to lose momentum, leading to an inward orbital decay.

\item As the subhalo orbits within the host system, the DM particles in the subhalo scatter with the DM particles in the host halo. These interactions inject energy into the subhalo, diffusing particles outward and forming a low-density cloud surrounding the subhalo, thereby lowering the central density. This extends the size of the core and can delay the process of core collapse. The SSHI process itself may directly contribute to the unbinding of some subhalo particles, reducing the bound mass; however, this effect does not play a dominant role in mass loss for the models we explore.
But, as the SSHI process reduces the gravitational binding of the DM subhalo particles, tidal stripping reduces the bound mass more easily.

Moreover, the SSHI process also leads to a drag force on the subhalo’s orbit \citep{Kummer:2017bhr}. Momentum exchange during host–subhalo scatterings leads to a net transfer of orbital energy from the subhalo to the host halo. As a result, the subhalo gradually loses orbital energy and angular momentum, causing its orbit to decay inwards.

It is important to note that the strength of the SSHI process is proportional to the cross section at orbital velocities (for scenarios where orbital velocity is much larger than the subhalo's velocity dispersion), while the heat transfer inside the subhalos is determined by the cross section at velocities of order the velocity disperion of the subhalo. The importance of the SSHI process is therefore directly related to the velocity dependence of the cross section.

\item Tidal forces accelerate the DM subhalo particles at all radii and are the strongest during pericentre passage. As a result, during a pericentre passage, there is a temporary increase in the velocity dispersion, and the high-velocity particles are tidally stripped: they become unbound and leave the system.
Tidal forces are the strongest at the outermost region of the subhalo; therefore, mass is primarily stripped from the outskirts, accelerating the gravothermal evolution of the subhalo's core \citep{Nishikawa_2020}. The stripped DM particles form leading and trailing DM streams, while the particles that remain gravitationally bound experience tidal heating: increase of the velocity dispersion due to energy injection from the tidal field. 

If tidal heating is strong, it can prevent or reduce the heat outflow from the centre of the halo, which can delay or reverse the process of core collapse \citep{Zeng_2022}.

\item Ultimately, the subhalo either undergoes core collapse (c) or is entirely disrupted (d) by tidal forces and the SSHI process for the set of SIDM models, initial conditions and orbits we simulated. 
\end{enumerate}

In this work, we focus on satellites with sufficiently small masses ($m_\mathrm{sub} \leq 1/1000 \ m_\mathrm{host}$) such that dynamical friction can be neglected \citep{Tormen:1997ik, Zeng_2022}.
The other environmental effects play a crucial role in subhalo evolution, and accurately modelling these processes is important to obtain robust predictions.

\section{Numerical method}
\label{sec:numerical_method}
In this section, we describe our numerical setup. We begin by describing the simulation code and its SIDM implementation. Then, we present our implementation of the SSHI process.
Finally, we discuss our procedures for extracting relevant halo properties from our simulations.

We use the cosmological hydrodynamical $N$-body code \textsc{OpenGadget3} (Dolag et al.\ in prep.), a successor of \textsc{Gadget-2} \citep{10.1111/j.1365-2966.2005.09655.x}. 
Gravity is treated through a one-sided oct-tree algorithm \citep{1986Natur.324..446B}, efficiently calculating the gravitational forces between numerical particles.
For more details, we refer the reader to the review by \citet{Dehnen_2011}, which provides a comprehensive overview of the underlying principles and implementation in $N$-body codes. To optimize particle interaction handling, the code uses advanced domain decomposition and neighbour search techniques \citep{ragagnin2016exploiting}. 

\subsection{Self-interacting dark matter}
$N$-body codes approximate the phase-space distribution of DM through a discrete set of numerical particles.
A numerical particle represents many physical particles, which are assumed to have the same velocity. The DM density distribution represented by a numerical particle $i$ with mass $m_i$ at position $\vec{x}_i$ is
\begin{equation}
    \rho_i(\vec{x})=m_i \, W(|\vec{x}-\vec{x_i}|, h_i) \,,
    \label{eq:DMDensityDistribution}
\end{equation}
where $W(|\vec{x}-\vec{x_i}|, h_i)$ is the kernel function, which depends on the distance to the numerical particle $|\vec{x}-\vec{x_i}|$ and its kernel size $h_i$.
The kernel is normalized such that its volume integral yields $1$.
We employ the common approach of setting the kernel size such that it includes a fixed number of neighbouring particles \citep{10.1111/j.1365-2966.2005.09655.x}, which we set to $N_\mathrm{ngb}=48$ for our simulations.
All numerical particles are assumed to have equal mass, $m_i = m$. 

The SIDM module of \textsc{OpenGagdet3} was developed by \citet{Fischer_2021a, Fischer_2021b, Fischer_2022, Fischer_2024, Fischer_2026b}. It assumes elastic scattering and contains two schemes: rare (rSIDM) and frequent self-interactions (fSIDM).
Self-interactions are implemented using a Monte Carlo scheme, where scattering events are probabilistically sampled based on pairs of numerical particles.
Such pairwise formulations are common among SIDM implementations in $N$-body codes \citep[e.g.][]{Koda:2011yb, Rocha:2012jg, Vogelsberger:2012ku, Fry:2015rta, Robertson:2016xjh, Correa_2022, Valdarnini_2024}. This module only considers pairs of particles that are close, i.e., for which the kernel functions overlap.

The rSIDM scheme models DM self-interactions in two steps: it determines if two particles scatter and then computes their post-scattering velocities. 
For each pair of particles $i$ and $j$ within the interaction radius, an interaction probability is calculated based on the self-interaction cross section per (fundamental particle) DM mass $\sigma/m_\chi$, relative velocity $\Delta\vec{v}_{ij}$, and time step $\Delta t$,
\begin{equation}
    P_{ij} = \frac{\sigma(|\Delta\vec{v_{ij}}|)}{m_\chi} \, m \, |\Delta\vec{v_{ij}}| \, \Delta t \, \Lambda_{ij} \,,
\end{equation}
where the kernel overlap $\Lambda_{ij} = \int W(|\vec{x}-\vec{x_i}|, h_i) \, W(|\vec{x}-\vec{x_j}|, h_j) \, \mathrm{d}^3 \mathbf{x}$ quantifies the spatial overlap of the particles’ density kernels and effectively captures the interaction volume. A pseudo-random number $y$ is drawn for each pair, and if $y \leq P_{ij}$, a scattering event occurs. To maintain accuracy, $\Delta t$ must be chosen such that $P_{ij} \ll 1$.

The scattering is performed in the centre-of-mass frame, with the outgoing directions drawn from the differential cross section. In our simulations, we use rSIDM to model isotropic scattering, but the method allows for arbitrary angular dependencies. However, for differential cross sections that dominate at small angles, rSIDM becomes computationally inefficient and prohibitively expensive due to the high number of interactions required \citep[e.g.][]{Arido_2025}.

The fSIDM module is tailored for the extreme forward-scattering regime. Instead of treating individual collisions, fSIDM approximates the cumulative effect of many weak scatterings as a continuous drag force with perpendicular momentum diffusion. The strength of this drag force depends on a modified momentum transfer cross section \citep{Kahlhoefer:2013dca}\footnote{We note that we adopt the normalisation as in \citep{Arido_2025}.},
\begin{equation} \label{eq:modified_momentum_transfer_cross_section}
    \sigma_\mathrm{\tilde{T}}=4 \uppi\int^1_{-1}\frac{\mathrm{d}\sigma}{\mathrm{d}\Omega_\mathrm{cms}}(1-|\cos\theta_\mathrm{cms}|)\,\mathrm{d}{ \cos{\theta_\mathrm{cms}}} \,.
\end{equation}
The drag force is
\begin{equation}
    F_\mathrm{drag}=\frac{1}{4}|\Delta \vec{v}_{ij}|^2\frac{\sigma_\mathrm{\tilde{T}}}{m_\chi}m^2\Lambda_{ij}
\end{equation}
and results in a velocity change $\Delta \vec{v}_\mathrm{drag} = F_\mathrm{drag} \Delta t / m$, symmetrically for both particles to conserve momentum and energy~\citep{Fischer_2021a, Fischer_2024}. The fSIDM module is not a general-purpose extension of rSIDM, but rather a dedicated scheme for modelling the extreme forward-scattering limit. The fSIDM module also provides a good approximation for other highly anisotropic cross sections \citep{Arido_2025}.

\subsection{Implementation of the SSHI process \label{ImplementationSSHIprocess}}

We model the gravitational potential of the host system analytically to avoid the high computational cost of resolving both the massive host and the much smaller satellite at high resolution within a single simulation. Our computational resources can then be dedicated to the satellite itself.

We also implement a dedicated procedure for the SSHI process, the scattering of DM particles in the subhalo with DM particles in the host.
We introduce virtual DM host particles to capture the SSHI process effect without explicitly populating the host halo with particles. To ensure physical accuracy, we construct the virtual particles to reproduce the local phase-space distribution of the host system. In other words, we sample host halo particles analogously to an initial conditions generator, but only locally, where and when needed.

At each time step, the simulation first processes the scatterings between DM subhalo particles, followed by our SSHI process procedure.
For each DM subhalo particle, we generate $N_\mathrm{virt}$ virtual DM particles of the host, sampled at the exact location of the DM subhalo particle. The existing SIDM module then simulates scattering between the DM subhalo particles and their virtual DM host particles.
These interactions are processed consecutively, and after each interaction, the velocity of the DM subhalo particle is updated. This approach effectively allows multiple scatterings per time step.
After the scattering calculations involving the virtual particles are complete, the virtual particles are discarded.

Our procedure ensures realistic scattering interactions between the DM particles in the subhalo and the DM particles in the host halo while keeping the simulation computationally efficient. However, the impact on the host is neglected.

To simulate SSHI process scattering with the SIDM module, it is necessary to calculate the kernel size $h$, which is needed for the kernel function in Eq.~\eqref{eq:DMDensityDistribution} and the kernel overlap $\Lambda_{ij}$ \citep[eq.~13 by][]{Fischer_2021a}. Here, we employ the same kernel size for both interaction partners, the virtual host particle and the subhalo particle, $h=h_i=h_j$. As both particles are at the exact same location ($d\equiv|\vec{x}_i - \vec{x}_j|=0$) and have the same kernel size, the kernel overlap simplifies to
\begin{equation}
    \Lambda_{ij}=\Lambda(h_i, h_j,d)=\Lambda(h,h,0)=\Lambda(1,1,0)/h^3 \equiv \Lambda_0 / h^3 \,.
\end{equation}
The kernel volume $h^3$ contains a representative number of virtual particles $N_\mathrm{virt}$, which have to match the local host density
\begin{equation}
    \rho_\mathrm{host}(\vec{x})=N_\mathrm{virt} \, \rho_\mathrm{virt}=N_\mathrm{virt} \, m \, \Lambda_0/h^3 \,.
    \label{eq:host_density}
\end{equation}
Thus, the SIDM module models scattering in a way that reflects the physical density of the host halo, capturing the SSHI process effect without explicitly simulating the entire host halo. From Eq.~\eqref{eq:host_density}, the kernel size is
\begin{equation}
    h=\left(\Lambda_0 \, m \, \frac{N_\mathrm{virt}}{\rho_\mathrm{host}(\vec{x})}\right)^{1/3} \,,
\end{equation}
where $m$ is the mass of the virtual particles, which we set to be equal to the mass of the subhalo particles. $\rho_\mathrm{host}(\vec{x})$ is the density of the host halo at the location of the subhalo particle, for which we sample the virtual particles. We set the number of virtual particles to be equal to the number of neighbouring particles of the SIDM module, $N_\mathrm{virt}=N_\mathrm{ngb}$.

We parallelise our SSHI process implementation for shared memory using open multi-processing (OpenMP). It does not require additional communication between processes via the Message Passing Interface (MPI). Subhalo particles are processed in parallel without restrictions, as they can be treated independently. However, for each subhalo particle, the associated virtual particles must be computed serially.

The virtual host particles should represent the local environment of the DM host halo to model the SSHI process scattering realistically. Therefore, they should also have typical velocities for the local host system.
Sampling with Eddington inversion reproduces a realistic and stable DM halo.
We implement the simplest case of Eddington inversion for a spherically symmetric and isotropic distribution. In Appendix~\ref{sec:verification_tests}, we describe the series of tests we have performed to verify our implementation of the SSHI process model.

\subsubsection{Comparison of our scheme to previous work}
\citet{Zeng_2022} developed the first quasi-analytic implementation of the SSHI process within the context of SIDM simulations, representing a significant step forward in modelling the environmental effects of SIDM on subhalo evolution.

Our approach builds upon and extends theirs in several important ways. First, while \citet{Zeng_2022} assumes the host halo particles follow a local Maxwell-Boltzmann velocity distribution, we go beyond this approximation. Using Eddington inversion allows us to construct a distribution function that naturally excludes the high-velocity tail associated with unbound particles (see Appendix~\ref{sec:VeriEddingtonInversion}). 
Second, their scheme is limited to only one scattering event per time step per particle. The implementation of SIDM in \textsc{OpenGadget3} supports multiple scattering events per particle within a single time step.
Third, their method is limited to isotropic cross sections, which limits its applicability to simple models and excludes, for example, light mediator models that feature far-forward scattering. Our implementation does not face this limitation. Since we model the DM self-interactions between particles of the subhalo and the host based on pairs of numerical particles, we can easily account for various angular and velocity dependencies.

\subsection{Extracting halo properties}

There are various features of the simulated halo/subhalo that we are interested in studying. Here, we describe our numerical prescription for obtaining these properties.

To obtain the density and radial size of the core of an SIDM subhalo, we first identify the centre of the subhalo by locating the minimum of the gravitational potential, using the code by \citet{Fischer_2021b}.
Based on this centre, the central density is calculated from the $10^3$ innermost particles.
The core radius $r_\mathrm{core}$ is then defined as the distance from the centre to the radius at which the density drops to half of the central density.

To determine the bound mass of the subhalo, we follow an iterative unbinding procedure.
Starting at the minimum of the gravitational potential, we iteratively identify and discard unbound particles. In each iteration, we compute the total energy of every particle with respect to the subhalo potential. 
We assume the subhalo moves with a velocity given by the average of all particles within a certain radius, which we take to be 0.12~kpc, around the potential minimum.
Particles with positive total energy are considered unbound and are excluded from the calculation of the potential in the next iteration. This process is repeated until no further particles are removed. The total bound mass is the sum of the masses of all remaining bound particles.
During this procedure, we neglect the contribution from the host to the gravitational potential and only consider the DM of the subhalo.

\section{SIDM halo evolution}
\label{sec:evolution}
In this section, we investigate the evolution of DM halos under the influence of elastic self-interactions.
We first study the case of a halo evolving in isolation (Sect.~\ref{sec:isolated_evolution}), followed by a satellite halo (Sect.~\ref{sec:satellite_evolution}).
Further, we consider two possibilities for the angular dependence of the cross section: isotropic (using the rSIDM submodule) and forward-dominated (using the fSIDM submodule).

\subsection{Isolated evolution}
\label{sec:isolated_evolution}
To reproduce the success of cosmological CDM simulations on large scales, SIDM models require a self-interaction cross section that becomes negligible at high relative velocities. Observations of galaxy clusters, which exhibit large velocity dispersions ($\sim 1000 \ \mathrm{km} \ \mathrm{s}^{-1}$), place stringent upper limits on the self-interaction cross section on the order of $\sigma/m_\chi \lesssim 0.1 \ \mathrm{cm}^2 \ \mathrm{g}^{-1}$ \citep{Andrade:2020lqq, Eckert:2022qia}.
At the same time, we are interested in large cross sections at low velocities to drive core collapse in low-mass subhalos, producing the high central densities observed in some satellite galaxies. If the cross section remains too large at high velocities, the SSHI process from host–subhalo interactions can prevent collapse \citep{Zeng_2022}. The interplay between efficient self-interactions in the subhalo and suppressed interactions with the host is essential for low-mass subhalos. Thus, the SIDM cross section needs to be large at low velocities and decrease at higher velocities \citep{Kaplinghat:2015aga}. We model the velocity-dependent cross section as \citep{Fischer_2024}
\begin{equation}
    \frac{\sigma_\mathrm{V}}{m_\chi}= \frac{\sigma_0}{m_\chi} \left[1+\left(\frac{v}{w}\right)^2\right]^{-2} \,,
    \label{eq:VelDependCross}
\end{equation}
where the constant prefactor $\sigma_0 / m_\chi$ and the velocity scale $w$ are parameters of the model. Depending on the angular dependency of the cross section (isotropic or forward-dominated), there are different normalization constants chosen so that all of the angle-averaged cross sections coincide \citep{Arido_2025}. For the isotropic case: $\sigma_\mathrm{tot}= \sigma_\mathrm{\tilde{T}}=\sigma_\mathrm{V}$ and for the forward-dominated case:
$\sigma_\mathrm{tot}\gg \sigma_\mathrm{\tilde{T}}=\frac{2}{3}\sigma_\mathrm{V}$.

We assume that the initial isolated halo has an NFW profile, characterised by a scale density $\rho_\mathrm{s}=\rho(r_\mathrm{s})$ and scale radius $r_\mathrm{s}$.
We choose a DM halo with a virial mass of $M_\mathrm{vir}=10^{10} \, \mathrm{M_{\odot}}$ and a high concentration $c_\mathrm{vir}=30$ from the mass-concentration relation~\citep{Dutton_2014}, corresponding to an NFW profile with $\rho_\mathrm{s} = 2.5 \times 10^{7} \, \mathrm{M_\odot \, \mathrm{kpc}^{-3}}$ and $r_\mathrm{s} =1.5 \, \mathrm{kpc}$. 

The core evolution of an isolated halo with a velocity-dependent self-interaction cross section can map onto the evolution for a velocity-independent cross section~\citep[e.g.][]{Yang_2022, Outmezguine_2023, Yang_2023}.
In order to facilitate such a mapping, we use the effective cross section~\citep{Yang_2022}
\begin{equation}
    \sigma_\mathrm{eff} = \frac{1}{512(\vdisp)^8} \
    \int v^5 \frac{2}{3} \ \sigma_V \ v^2 \ \exp\left[-\frac{v^2}{4(\vdisp)^2}\right] \ \mathrm{d}v \,,
    \label{eq:EffCrossSection}
\end{equation}
where we assume a Maxwell-Boltzmann distribution for the relative velocity of DM particles within the halo, and the characteristic 1D velocity dispersion is estimated to be $\vdisp = 1.1\, v_\mathrm{max}/\sqrt{3}$. Here, $v$ denotes the relative velocity in a given system, while $v_\mathrm{max}$ is the maximum of the circular velocity $v_\mathrm{circ}(r)=\sqrt{\mathrm{G} M(<r)/r}$ of the halo at a corresponding radius $r_\mathrm{max}$. The quantities $v_\mathrm{max}$ and $r_\mathrm{max}$ are key observables; they are connected to the density structure of DM halos and can be derived from the observed line-of-sight velocity dispersion. Note that for the case of a constant cross section, $\sigma_\mathrm{eff}$ coincides with $\sigma = \sigma_0$.

\begin{figure*}
    \centering
    \includegraphics[width=\columnwidth]{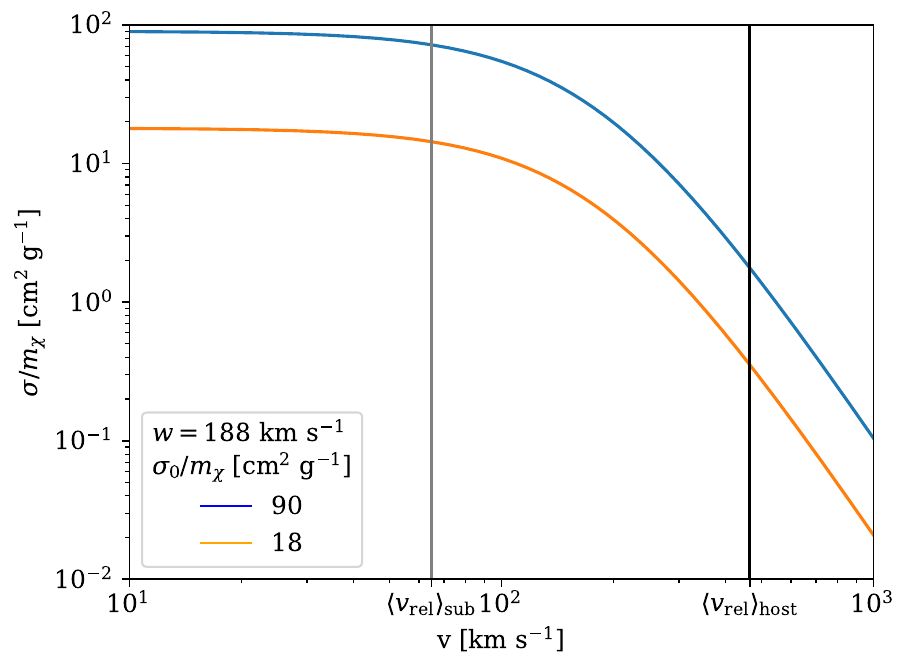}
    \includegraphics[width=\columnwidth]{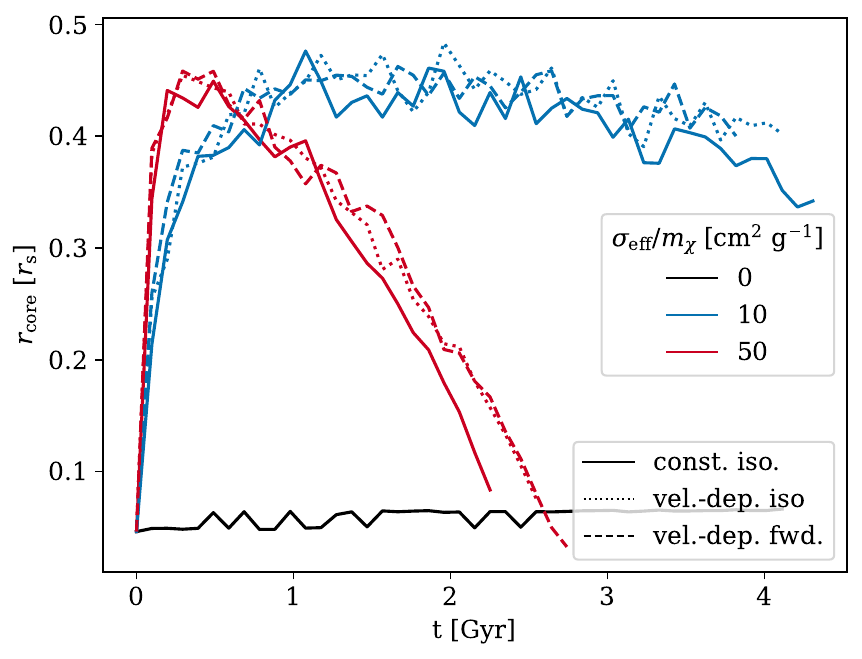}
    \caption{Left panel: The adopted velocity-dependent cross sections are shown together with the typical relative velocities $\langle v_\mathrm{rel}\rangle$ within the host (black) and subhalo (grey). Right panel: Evolution of the core size $r_\mathrm{core}$ of an isolated halo with three different effective cross sections. The radius $r_\mathrm{core}$ is given in units of the scale radius of the initial NFW profile. For each effective cross section, we simulate a constant isotropic cross section (solid), a velocity-dependent isotropic cross section (dotted), and a velocity-dependent forward-dominated cross section (dashed).}
    \label{fig:Cross-section}
\end{figure*}

We consider two velocity-dependent self-interaction cross sections, calibrated to map onto the core evolution of isolated SIDM halos with two different constant cross sections. These models have the general feature of enhanced cross section at low relative velocities, which declines at high relative velocities as expected of viable SIDM models that increase the diversity of DM mass profiles in galaxies~\citep{Kaplinghat:2015aga}. 
We choose [$\sigma_0/m_\chi=18 \ \mathrm{cm}^2 \ \mathrm{g}^{-1}$, $w=188 \ \mathrm{km} \ \mathrm{s}^{-1}$] and [$\sigma_0/m_\chi=90 \ \mathrm{cm}^2 \ \mathrm{g}^{-1}$, $w=188 \ \mathrm{km} \ \mathrm{s}^{-1}$] for the velocity-dependent cross sections corresponding to effective cross sections ($\sigma_\mathrm{eff}$) of [$10 \ \mathrm{cm}^2 \ \mathrm{g}^{-1}$, $50 \ \mathrm{cm}^2 \ \mathrm{g}^{-1}$], respectively.
The value of $w$ controls the impact of SSHI process; smaller values of $w$ would make the SSHI process irrelevant. The values of $\sigma_0/m_\chi$ and $w$ we chose are  consistent with the constraints on group and cluster scales~\citep{Sagunski:2020spe,Andrade:2020lqq} and may lead to core collapse in low-mass halos~\citep{Nishikawa_2020,Roberts:2024uyw}.

In the left panel of Fig.~\ref{fig:Cross-section}, we plot the velocity-dependent cross sections that we employ in our simulations. 
The velocity-dependent cross sections are shown as a function of relative velocities for the scattering. Average relative velocities of the subhalo and host halo are shown as vertical lines and are estimated via $\langle v_\mathrm{rel}\rangle=\int^\infty_0v_\mathrm{rel} \, f(v_\mathrm{rel}) \, dv_\mathrm{rel}= 4 \, \vdisp/\sqrt{\uppi} $ by using the Maxwell-Boltzmann distribution $f(v)$. The host halo is described in the next section (Sect.~\ref{sec:satellite_evolution}).

To validate the mapping for the evolution of an isolated DM halo with velocity-dependent and constant cross sections, we perform simulations of an isolated DM halo with a constant isotropic cross section and two simulations [one with isotropic (simulated as rSIDM) and one with forward-dominated (simulated as fSIDM) scattering] with the corresponding velocity-dependent cross section. The effective cross sections for the forward-dominated scattering and isotropic scattering are the same. The DM halo is resolved with $N=3 \times 10^6$ particles. These are sampled using the \textsc{SpherIC} code~\citep{Garrison-Kimmel:2013yys}, resulting in a numerical particle mass of $m=277 \ \mathrm{M}_\odot$. The simulations use a gravitational softening length of $\epsilon = 0.03 \ \mathrm{kpc}$ and a neighbour number of $N_\mathrm{ngb} = 48$. We set a spherical boundary at a radius of $6000 \ \mathrm{kpc}$, at which the particles are reflected, in order to prevent significant round-off errors from particles at large distances \citep{Fischer_2024b}.

The right panel of Fig.~\ref{fig:Cross-section} shows the evolution of the core size for the isolated halos.
We find that for the simulations corresponding to a given effective cross section, the core evolution is consistent within numerical inaccuracies. As core evolution progresses in the core collapse phase, the difference between the constant and velocity-dependent cross sections increases, consistent with expectations from gravothermal studies~\citep{Gad-Nasr:2023gvf}. The larger the velocity scale $w$ is compared to the scattering velocities, the better the matching works. The slower collapse of the velocity-dependent cross section is in line with previous findings~\citep{Fischer_2024, Fischer_2025}.

\subsection{Satellite evolution}
\label{sec:satellite_evolution}
The goal of this section is to understand how SIDM subhalos evolve when orbiting within a host halo and how the interplay of self-interactions, tidal forces, and the SSHI process influences their structure over time. In contrast to the isolated case studied in Sect.~\ref{sec:isolated_evolution}, we now investigate the impact of an external environment on the evolution of velocity-dependent SIDM subhalos. We examine the difference between the CDM and SIDM subhalos, as well as the difference between isotropic and forward-dominated self-interactions.
\definecolor{mycolor1}{HTML}{1b9e77}
\definecolor{mycolor2}{HTML}{d95f02}
\definecolor{mycolor3}{HTML}{7570b3}

\newcommand{\fullcircle}{\tikz \fill (0,0) circle (0.8ex);}

\newcommand{\upperhalfcircle}{\tikz \fill (0,0) arc (0:180:0.8ex);}

\newcommand{\lowerhalfcircle}{\tikz \fill (0,0) arc (180:360:0.8ex);}

\newcommand{\lefthalfcircle}{\tikz \fill (0,0) -- ++(0,0.8ex) arc (90:270:0.8ex) -- cycle;}

\newcommand{\righthalfcircle}{\tikz \fill (0,0) -- ++(0,-0.8ex) arc (-90:90:0.8ex) -- cycle;}

\newcommand{\fulldiamond}{\tikz \fill (0,0.8ex) -- (0.8ex,0) -- (0,-0.8ex) -- (-0.8ex,0) -- cycle;}

\newcommand{\upperhalfdiamond}{\tikz \fill (0,0.8ex) -- (0.8ex,0) -- (-0.8ex,0) -- cycle;}

\newcommand{\lowerhalfdiamond}{\tikz \fill (0,-0.8ex) -- (0.8ex,0) -- (-0.8ex,0) -- cycle;}

\newcommand{\lefthalfdiamond}{\tikz \fill (-0.8ex,0) -- (0,0.8ex) -- (0, -0.8ex) -- cycle;}

\newcommand{\righthalfdiamond}{\tikz \fill (0.8ex,0) -- (0,0.8ex) -- (0, -0.8ex) -- cycle;}

\newcommand{\fullstar}{\tikz \fill[scale=0.008] (0,25) -- (7,7) -- (25,7) -- (10,-3) -- (15,-20) -- (0,-8) -- (-15,-20) -- (-10,-3) -- (-25,7) -- (-7,7) -- cycle;}

\newcommand{\upperhalfstar}{\tikz \fill[scale=0.008] (0,25) -- (7,7) -- (25,7) -- (10,-3) -- (0,-3) -- (-10,-3) -- (-25,7) -- (-7,7) -- cycle;}

\newcommand{\lowerhalfstar}{\tikz \fill[scale=0.008] (0,-8) -- (-15,-20) -- (-10,-3) -- (0,-3) -- (10,-3) -- (15,-20) -- cycle;}

\newcommand{\lefthalfstar}{\tikz \fill[scale=0.008]
  (0,25) -- (-7,7) -- (-25,7) -- (-10,-3) -- (-15,-20) -- (0,-8) -- (0,25) -- cycle;}

\newcommand{\righthalfstar}{\tikz \fill[scale=0.008]
  (0,25) -- (7,7) -- (25,7) -- (10,-3) -- (15,-20) -- (0,-8) -- (0,25) -- cycle;}

\begin{table}
    \caption{Orbital and cross section parameters.}
    \centering
    \begin{tabular}{lccccc}
        \hline\hline
        \textbf{DM Model ($\sigma_\mathrm{eff}/m_\chi$)} 
        & eo=2--1 
        & eo=5--1 
        & eo=6--3 \\
        \hline
        CDM ($0 \ \mathrm{cm}^2 \ \mathrm{g}^{-1}$) 
        & {\Large\color{mycolor1} \fullcircle} 
        & {\Large\color{mycolor2} \fullstar} 
        & {\Large\color{mycolor3} \fulldiamond} \\
        
        rSIDM ($50 \ \mathrm{cm}^2 \ \mathrm{g}^{-1}$) 
        & {\Large\color{mycolor1} \upperhalfcircle} 
        & {\Large\color{mycolor2} \upperhalfstar} 
        & {\Large\color{mycolor3} \upperhalfdiamond} \\
        
        rSIDM ($10 \ \mathrm{cm}^2 \ \mathrm{g}^{-1}$) 
        & {\Large\color{mycolor1} \lowerhalfcircle} 
        & {\Large\color{mycolor2} \lowerhalfstar} 
        & {\Large\color{mycolor3} \lowerhalfdiamond} \\
        
        fSIDM ($50 \ \mathrm{cm}^2 \ \mathrm{g}^{-1}$) 
        & {\Large\color{mycolor1} \righthalfcircle} 
        & {\Large\color{mycolor2} \righthalfstar} 
        & {\Large\color{mycolor3} \righthalfdiamond} \\
        
        fSIDM ($10 \ \mathrm{cm}^2 \ \mathrm{g}^{-1}$) 
        & {\Large\color{mycolor1} \lefthalfcircle} 
        & {\Large\color{mycolor2} \lefthalfstar} 
        & {\Large\color{mycolor3} \lefthalfdiamond} \\
        \hline
    \end{tabular}
    \tablefoot{Simulations for CDM and four SIDM scenarios are performed for three different elliptical orbits (eo). For SIDM, the two velocity-dependent cross sections from Sect.~\ref{sec:isolated_evolution} are considered; each is implemented with both isotropic (rSIDM) and forward-dominated scattering (fSIDM). The notation $\mathrm{eo}=\mathrm{A}$--$\mathrm{B}$ denotes the apocentre (A) and pericentre (B) distances, given in units of the host's scale radius, $r_\mathrm{s, \, host}$. All cross sections are given in terms of the effective cross section as defined in Eq.~\eqref{eq:EffCrossSection}, following \citet{Yang_2022}.}
    \label{tab:orbits}
\end{table}

The satellite galaxy is represented by a DM subhalo, which has the same halo properties as the isolated DM halo in Sect.~\ref{sec:isolated_evolution}. This highly concentrated DM subhalo has a typical mass scale for dwarf satellite galaxies.
The host halo is analytically modelled and has a mass of $M_\mathrm{vir} = 10^{13} \ \mathrm{M_\odot}$ and a typical concentration $c_\mathrm{vir} = 8.8$ of the mass-concentration relation \citep{Dutton_2014}, with an NFW profile defined by $\rho_\mathrm{s, \, host} = 1.1 \times 10^6 \ \mathrm{M_\odot} \ \mathrm{kpc}^{-3}$ and $r_\mathrm{s, \, host} = 51 \ \mathrm{kpc}$.
We select three characteristic orbits to qualitatively analyse the evolution of satellite galaxies. For each orbit, we perform a CDM simulation and SIDM simulations for two velocity-dependent cross sections, each with a different choice of the angular distribution of scattering. The parameters of the different simulations are given in Table~\ref{tab:orbits}.

In the right panel of Fig.~\ref{fig:SubhaloEvolution}, the density maps are shown for the DM subhalos on the inner orbit (eo=2--1) for CDM (top row), isotropic SIDM without the SSHI process (middle row), and isotropic SIDM with the SSHI process (bottom row).
The CDM subhalo retains its cuspy central density profile, while tidal forces strip particles from their outskirts, forming DM streams that lead and trail the subhalo.
The SIDM subhalo without the SSHI process develops a core, a central region of approximately constant density, due to self-interactions. This core undergoes collapse in the final stages of the simulation.
In contrast, the SIDM subhalo with the SSHI process also forms a core early on but reaches a larger maximum core size. The SSHI process primarily affects the dense central region of the subhalo, reducing the central density of the subhalo, thereby suppressing the typical central density increase associated with core collapse.

\begin{figure}
    \centering
    \includegraphics[width=\columnwidth]{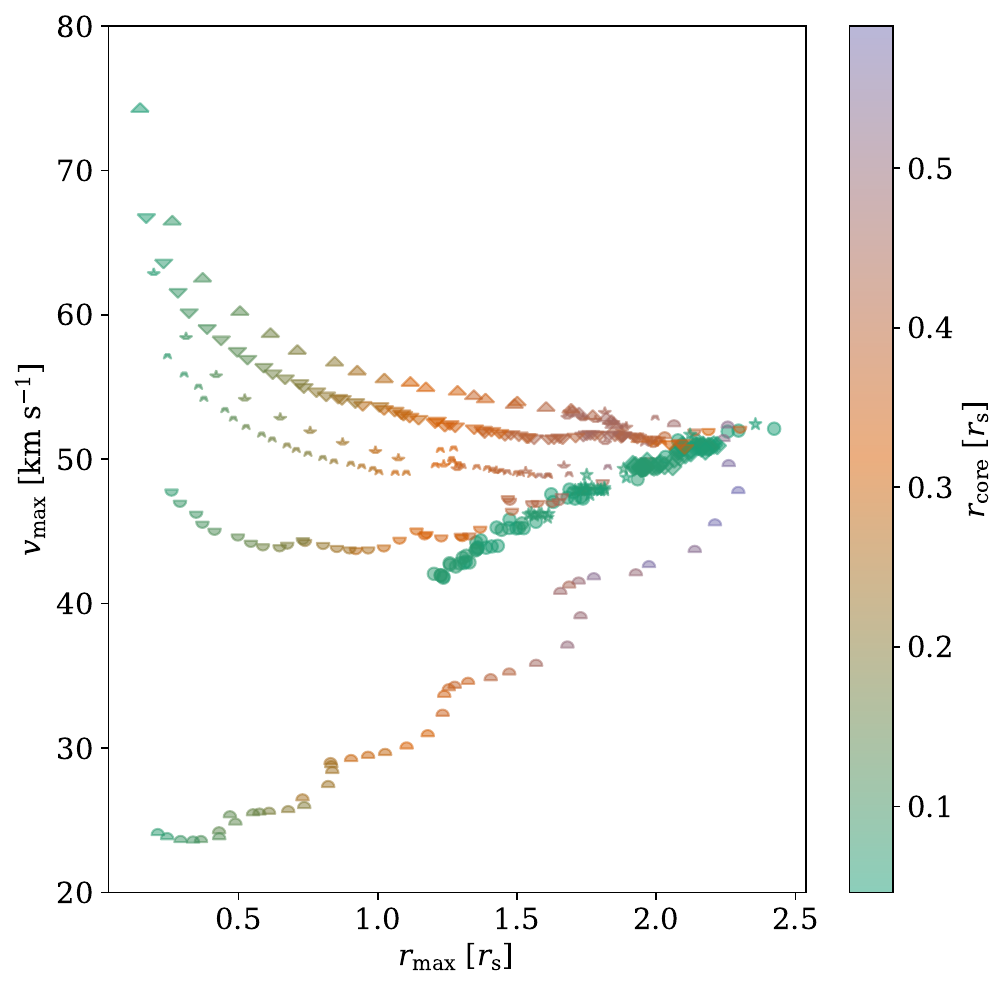}
    \caption{Evolution of the maximum circular velocity $v_\mathrm{max}$ and its corresponding radius $r_\mathrm{max}$. The results for various orbits and cross sections are displayed using the symbols as given in Table~\ref{tab:orbits}. The colour indicates the DM core size, $r_\mathrm{core}$. The radii $r_\mathrm{max}$ and $r_\mathrm{core}$ are given in units of the scale radius of the initial NFW subhalo.}
    \label{fig:VMAX}
\end{figure}

In Fig.~\ref{fig:VMAX}, the maximum circular velocity and the corresponding radius are shown for the CDM and isotropic SIDM subhalos as they evolve with time. The colour bar indicates the core size at each time. The CDM subhalos always have a green colour as they do not form a core.
Comparing subhalos along different orbits, there is substantially more variation in $v_\mathrm{max}$ for the SIDM subhalos than the CDM subhalos, and the variation is larger for the simulations with the larger SIDM cross section.

\begin{figure}
    \centering
    \includegraphics[width=\columnwidth]{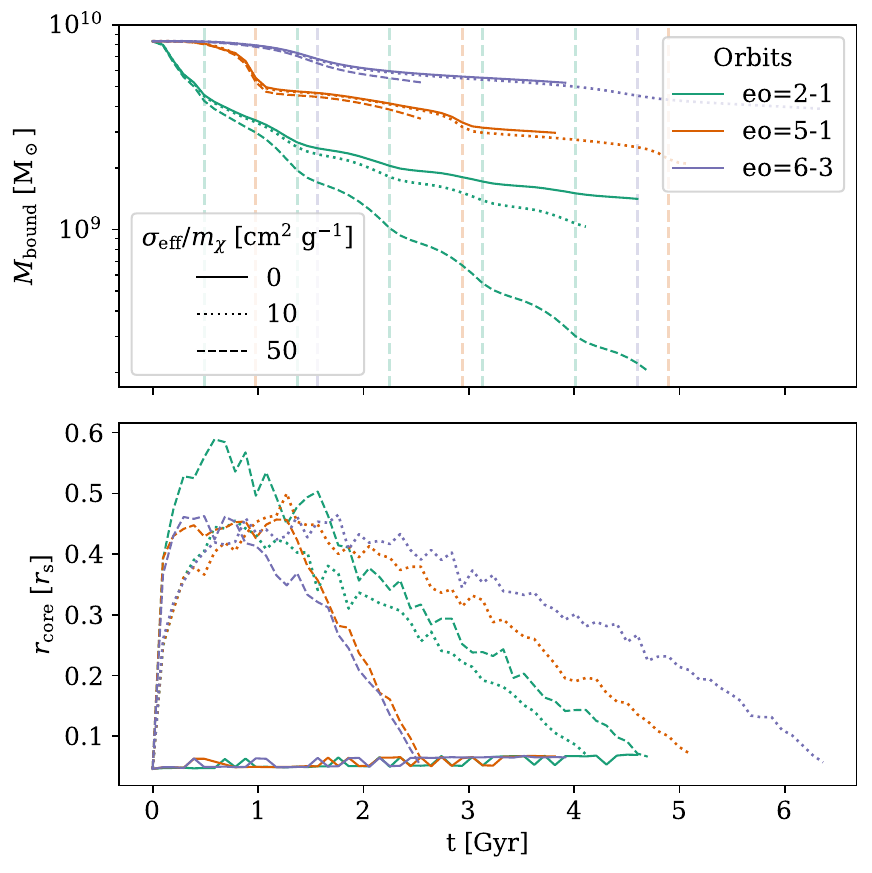}
    \caption{Bound mass $M_\mathrm{bound}$ (top panel) and core size $r_\mathrm{core}$ (bottom panel) as a function of time. The radius $r_\mathrm{core}$ is given in units of the scale radius of the initial NFW subhalo. The three different orbits are distinguished by colour, and the corresponding pericentres are indicated by the dashed vertical lines. Each orbit is shown for CDM (solid) and for SIDM with isotropic scattering for two different cross sections (dotted and dashed).}
    \label{fig:Evolution}
\end{figure}

For CDM subhalos, the evolution of $v_\mathrm{max}$ and $r_\mathrm{max}$ is linear, decreasing with time on well-defined tidal tracks \citep[e.g.][]{St_cker_2023, Du_2024}. 
In contrast, SIDM subhalos follow a more complex, non-linear evolution \citep{ zeng2025diversityuniversalityevolutiondwarf}. For SIDM subhalos, the maximum circular velocity slowly increases but accelerates over time, especially during core collapse, while the corresponding radius decreases with time.
The exception is the SIDM subhalo ({\Large\color{mycolor1} \upperhalfcircle}), for which tidal forces and the SSHI process decrease the maximum circular velocity almost until the end of our simulation.
This effect is most significant during pericentre approaches: there is a substantial drop in the maximum circular velocity, and the corresponding radius increases slightly or remains constant. This is connected to the significant mass loss of this subhalo (see Fig.~\ref{fig:Evolution} dashed green line).

In Fig.~\ref{fig:Evolution}, the top panel shows the evolution of the bound mass $M_\mathrm{bound}$ for DM subhalos on different orbits. The vertical lines in the bottom panel indicate the pericentre.
Subhalos in closer orbits experience stronger tidal forces and therefore greater mass loss. In SIDM, the formation of a density core leads to a more weakly bound structure compared to CDM. Thus, SIDM subhalos are more susceptible to tidal stripping, especially during pericentre passages, which is reflected in the more rapid decline of the bound mass.

Tidal stripping tends to remove high-energy particles. As a core forms with SIDM, the gravitational potential becomes shallower, increasing the phase-space volume occupied by weakly bound particles and enhancing the efficiency of stripping. Importantly, the self-interactions continuously scatter particles and can repopulate the high-velocity tail of the velocity distribution. This sustained replenishment of loosely bound particles may further facilitate mass loss by maintaining a reservoir of particles susceptible to tidal disruption.

The SIDM subhalos with the higher effective cross section experience greater mass loss than those with lower effective cross sections. Therefore, the contrast between CDM and SIDM subhalos grows with increasing effective cross section. However, subhalos with higher cross sections tend to form cores more rapidly, resulting in an earlier deviation in mass compared to that of CDM. In contrast, subhalos with lower cross sections often maintain their maximum core size for a longer period, allowing for extended mass loss over time.

The bottom panel of Fig.~\ref{fig:Evolution} shows the different evolution of the core size $r_\mathrm{core}$. For SIDM subhalos on wide orbits, the evolution closely resembles that of an isolated halo (see the right panel of Fig.~\ref{fig:Cross-section}). However, as the satellite moves closer to the centre of the host halo, the influence of the host environment becomes stronger. The core evolution diverges significantly from that of the isolated halo, particularly near the pericentre, where environmental effects are strongest. To isolate the impact of the SSHI process, we also ran the higher SIDM cross section simulation without the SSHI process; see Sec.~\ref{sec:fSIDM}. During the pericentre passage, tidal heating alone causes a slight increase in the maximum core size compared to an isolated halo, and the core begins to collapse shortly afterwards. In contrast, when the SSHI process is included, the core evolution is significantly prolonged.

The pericentre passage also marks a period of intensified mass loss due to tidal stripping and increased effect of the SSHI process. The reduction in the gravitational binding of the subhalo makes it even more susceptible to environmental effects. As mass is lost primarily from the outskirts, the velocity dispersion profile steepens in the outer regions, which enhances the outward heat flow and accelerates the onset of core collapse.
The resulting core evolution is dependent on the interplay between energy injection (from tidal heating and the SSHI process) and weaker gravitational binding (from tidal stripping and the SSHI process), making pericentre a critical point in the subhalo’s evolutionary history.

\begin{figure*}
    \centering
    \includegraphics[width=\columnwidth]{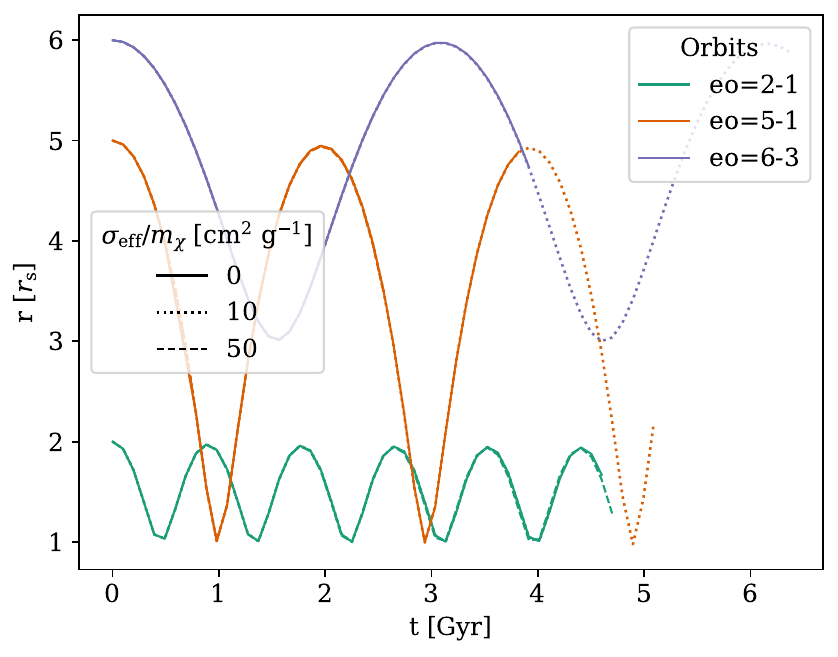}
    \includegraphics[width=\columnwidth]{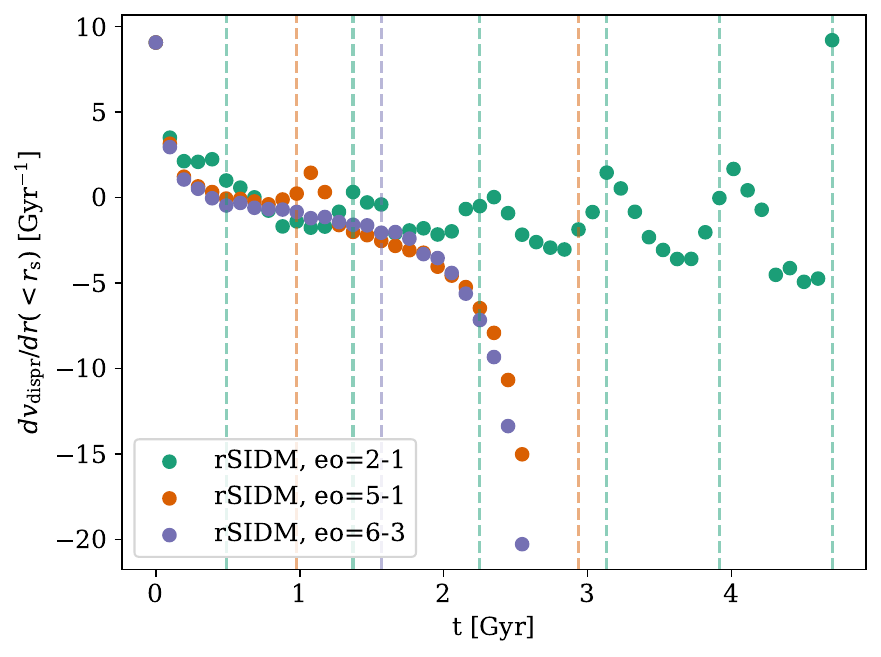}
    \caption{The left panel shows the distance to the centre of the host halo. The radius $r$ is given in units of the scale radius of the initial NFW host halo. The right panel shows the gradient of the velocity dispersion $\mathrm{d} v_\mathrm{dispr}/\mathrm{d}r$ within the initial $r_\mathrm{s}$ of the subhalos for the higher cross section $\sigma_\mathrm{eff}/m_\chi = 50 \  \mathrm{cm}^2 \ \mathrm{g}^{-1}$. The dashed vertical lines indicate the corresponding pericentres.}
    \label{fig:HeatFlow}
\end{figure*}

The distances to the host centre for the different orbits are shown in the left panel of Fig.~\ref{fig:HeatFlow}. There is no significant difference in the orbital decay between CDM and isotropic SIDM. The right panel shows the gradient for the velocity dispersion $\mathrm{d} v_\mathrm{dispr}/\mathrm{d}r$ in the central region ($r < r_\mathrm{s}$). This gradient is proportional to the temperature gradient, which indicates if the subhalo is in the core expansion ($\mathrm{d} v_\mathrm{dispr}/\mathrm{d}r>0$) or collapse phase ($\mathrm{d} v_\mathrm{dispr}/\mathrm{d}r<0$) by indicating the direction of the heat flow (inward or outward).

The SIDM subhalo on the outer orbit (purple) shows an evolution of the gradient that should be very similar to that of an isolated halo.
For the next inner orbit (orange), the gradient suddenly increases and becomes positive again around $1 \ \mathrm{Gyr}$, when the subhalo approaches its pericentre. Environmental effects suppress core collapse, initiating a core expansion phase. As a result, there is a more extended phase of the maximum core size, as seen in the bottom panel of Fig.~\ref{fig:Evolution} by comparing the dashed purple and dashed orange lines. During core collapse, the dashed orange line catches up with the dashed purple line because the higher mass loss accelerates the core evolution. The same behaviour is seen in Fig.~\ref{fig:HeatFlow}.

In Fig.~\ref{fig:HeatFlow}, for the SIDM subhalo on the inner orbit (green), every pericentre passage initiates a core expansion phase. The resulting subhalo has a large maximum core size and strongly extended core evolution. If the velocity dependence of the cross section is lower (higher velocity scale $w$, see Eq.~\eqref{eq:VelDependCross}), the effects of the SSHI process and tidal heating increase, potentially preventing the core collapse, which results in the likely destruction of the SIDM subhalo.

\subsubsection{Tidal heating}
\label{sec:TidalHeating}
\begin{figure}
    \centering
    \includegraphics[width=\columnwidth]{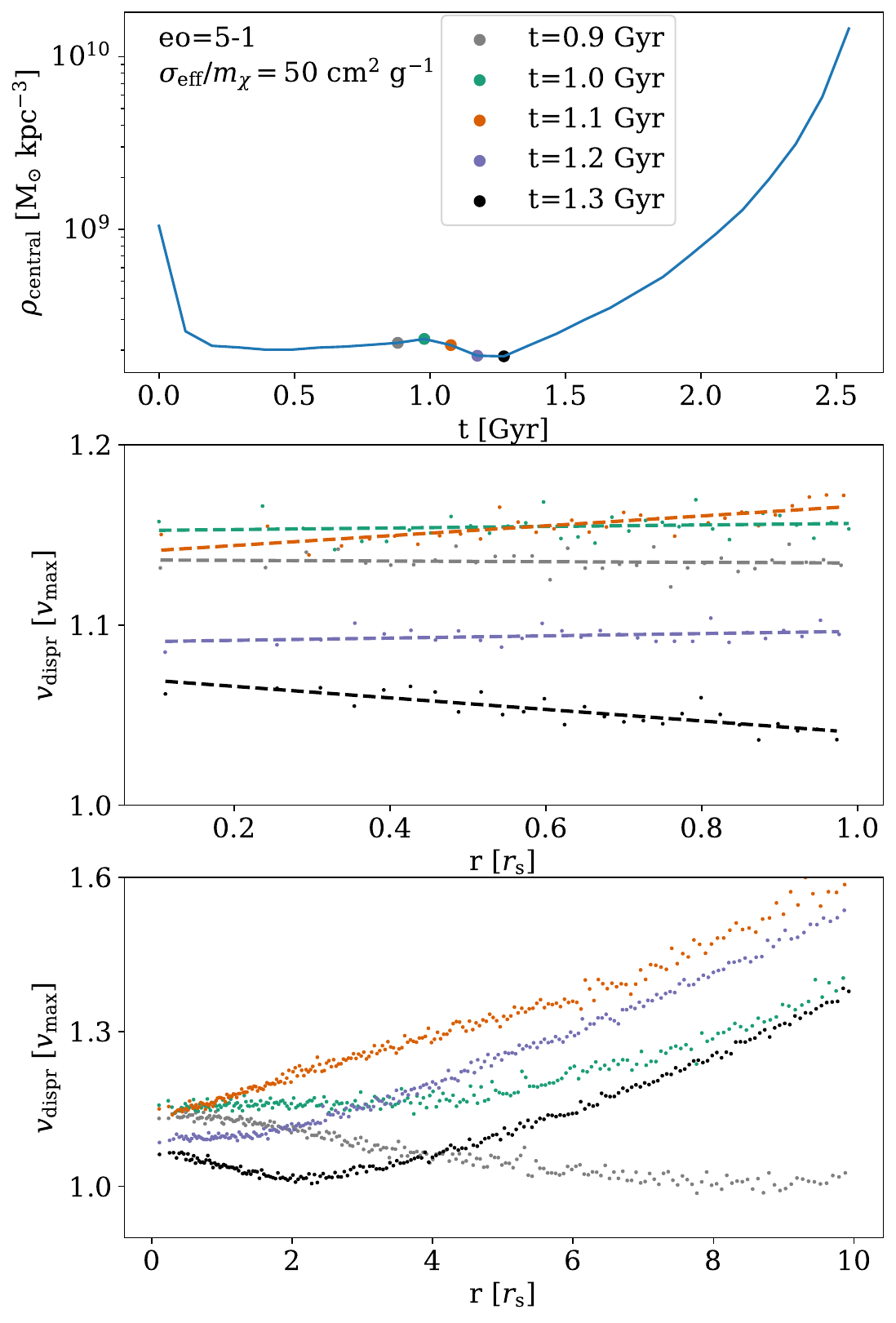}
    \caption{This figure highlights the effect of tidal heating on the SIDM subhalo with $\sigma_\mathrm{eff}/m_\chi = 50 \ \mathrm{cm}^2 \ \mathrm{g}^{-1}$ and eo=5--1. The top panel illustrates the evolution of the central density as a blue line. The middle panel shows the central velocity dispersion profile. The bottom panel shows the full radial velocity dispersion profile. $v_\mathrm{max}$ is here the initial value of the NFW profile. The radius $r$ is given in units of the scale radius of the initial NFW subhalo.}
    \label{fig:TidalHeating}
\end{figure}

To understand how tidal heating influences subhalo evolution, we consider the subhalo on the intermediate orbit ({\Large\color{mycolor2} \upperhalfstar}) around its pericentre. This subhalo is also subject to the SSHI process. To assess the impact of the SSHI process, we compared the results to a simulation without the SSHI process, but found no significant difference in the tidal heating effect around pericentre. Figure~\ref{fig:TidalHeating} illustrates how tidal heating influences the evolution of the SIDM subhalo with the higher cross section of $\sigma_\mathrm{eff} / m_\chi = 50 \ \mathrm{cm}^2 \ \mathrm{g}^{-1}$ on the intermediate orbit (eo=5--1). The top panel shows the evolution of the central density. 

At $t = 0.9 \, \mathrm{Gyr}$, the subhalo is in the core collapse phase, evident from the increasing central density in the top panel and the constant central velocity dispersion in the middle panel of Fig.~\ref{fig:TidalHeating}. 
During core collapse, both the central density and the velocity dispersion increase between $t = 0.9 \, \mathrm{Gyr}$ and $t = 1.0 \, \mathrm{Gyr}$, as shown in the top and middle panels.

At $t = 1.0 \, \mathrm{Gyr}$, the subhalo reaches its pericentre, where environmental effects from the host halo are strongest. The impact is visible in the bottom panel: particles in the outer region are significantly accelerated, increasing the velocity dispersion in the outer regions of the subhalo, which exceeds the velocity dispersion in the core. Self-interactions transport heat from the outer to the inner areas, inducing a core expansion phase, reflected in the drop in central density at $t = 1.1 \, \mathrm{Gyr}$ in the top panel and the positive temperature gradient in the middle panel. Core expansion continues through $t = 1.2 \, \mathrm{Gyr}$. 

By $t = 1.3 \, \mathrm{Gyr}$, the subhalo has moved sufficiently far from the pericentre for the influence of tidal heating to largely cease. In the middle panel of Fig.~\ref{fig:TidalHeating}, the temperature gradient once again becomes negative in the inner region, signalling the re-initiation of the core collapse phase. The transition back to core collapse occurs rapidly, highlighting that the periods of induced core expansion are short-lived and quickly dominated by core collapse. 

In Appendix~\ref{sec:DensityMassPofiles} additional information is provided to highlight how tidal heating influences the velocity dispersion of the SIDM subhalo. There, we show the evolution of the velocity dispersion within the core, as well as the density and velocity dispersion maps around the pericentre.

\subsubsection{Projected density slope and mass}
\begin{figure*}
    \centering
    \includegraphics[width=\columnwidth]{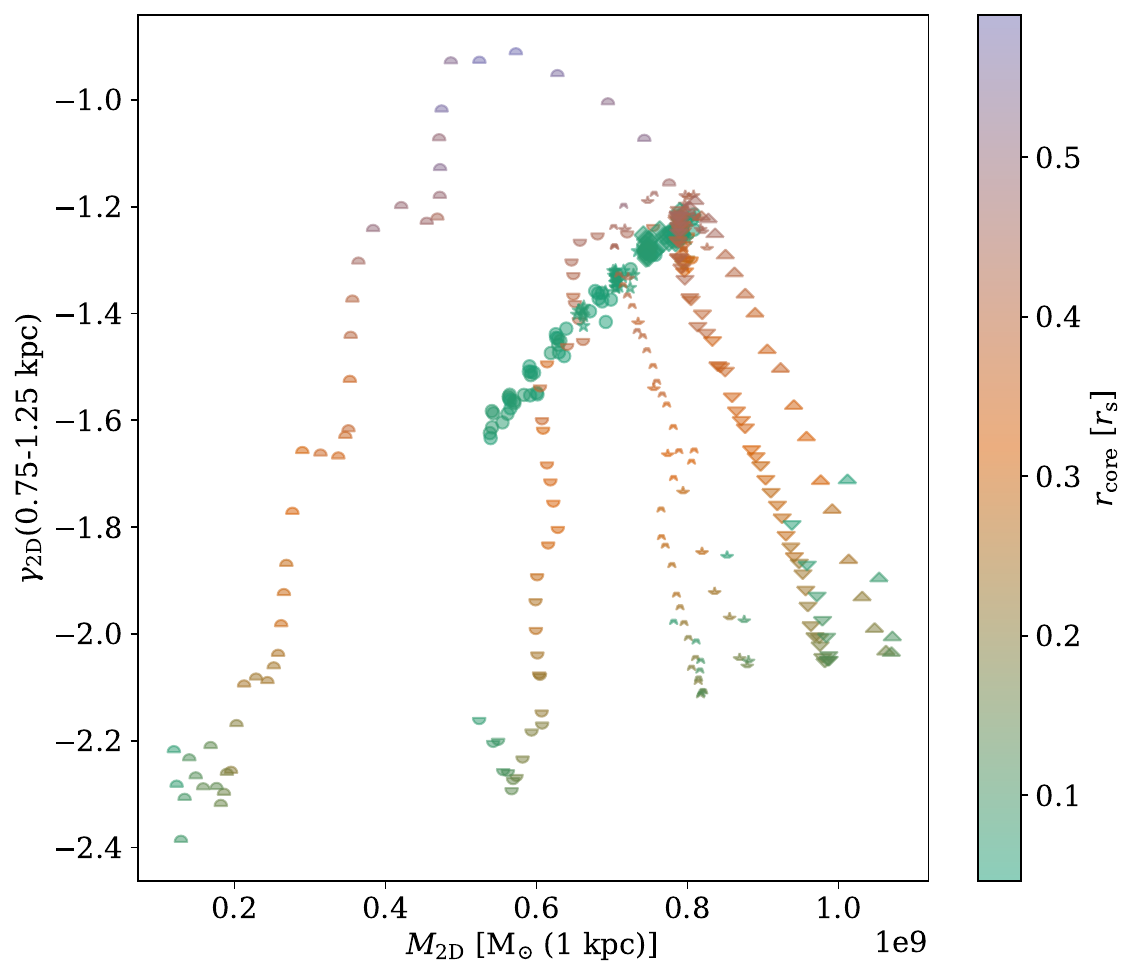}
    \includegraphics[width=\columnwidth]{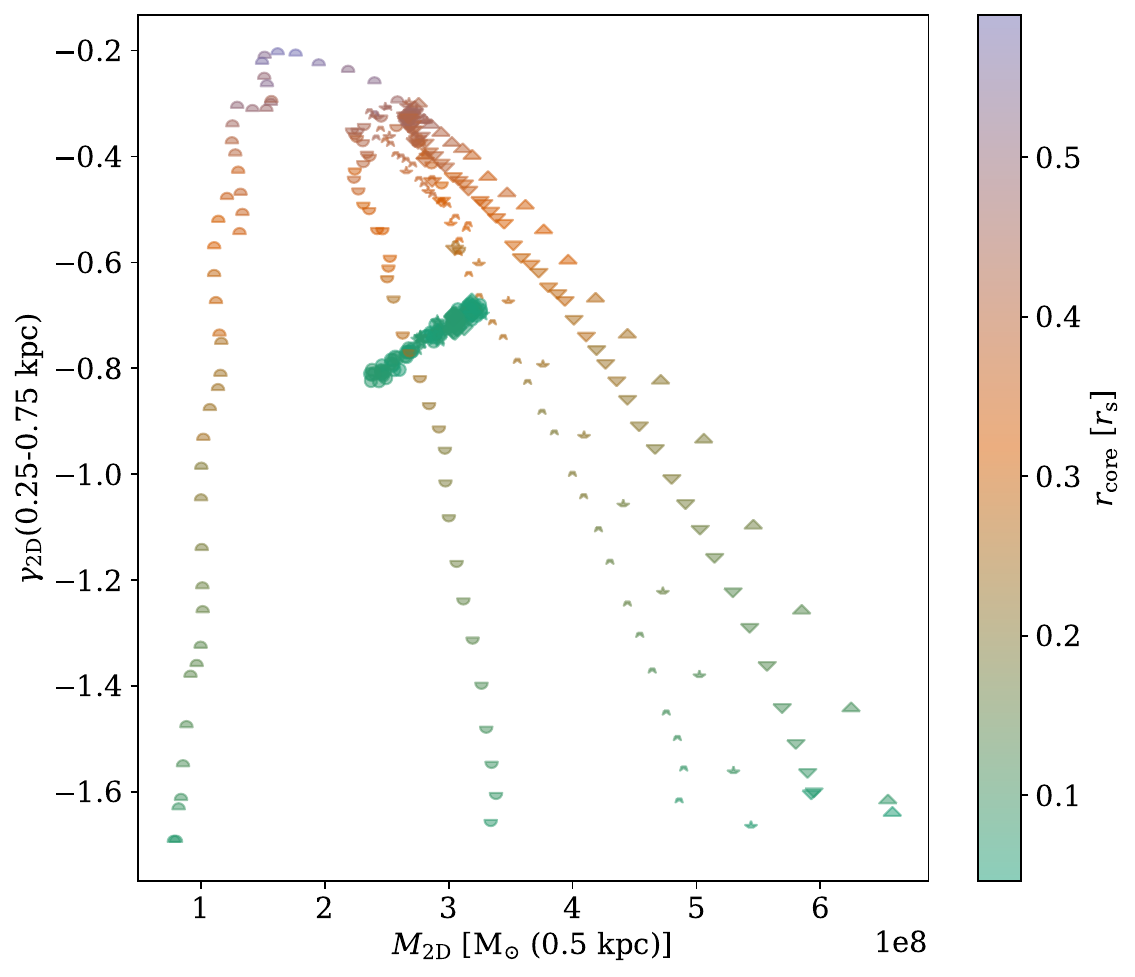}
    \caption{Left panel shows the logarithmic projected density slope $\gamma_\mathrm{2D}$ between 0.75 kpc and 1.25 kpc versus the projected mass $M_\mathrm{2D}$ at 1 kpc, and the right panel shows the evolution of the logarithmic projected density slope between 0.25 kpc and 0.75 kpc versus the projected mass at 0.5 kpc. The simulation results for various orbits and cross sections are displayed using the symbols as given in the Table~\ref{tab:orbits}. The colour indicates the DM core size, $r_\mathrm{core}$, in units of the scale radius of the initial NFW subhalo. In both panels, the SIDM subhalo evolution tracks the decreasing core size (from purple to green); the CDM halos evolve nearly linearly toward smaller values of $\gamma_\mathrm{2D}$ and $M_\mathrm{2D}$ (remaining green, which reflects their persistently small ``core'' size).}
    \label{fig:ProjectedMass/Slope}
\end{figure*}
The projected density slope and projected mass are key quantities for comparing DM subhalos in simulations with those inferred from observations of disturbed gravitational lenses. The projected mass is the enclosed mass within a given projected radius; the projected density slope is the logarithmic radial gradient of the density, obtained by integrating along the line of sight from a chosen viewing direction. Figure~\ref{fig:ProjectedMass/Slope} shows the projected density slope versus the projected mass for our CDM and rSIDM subhalos. The quantities are calculated from a face-on view, perpendicular to the orbital plane. For the left panel, the projected density slope is calculated as the average within a radial range of $r_\mathrm{2D} = 0.75 \, $--$\, 1.25 \, \mathrm{kpc}$, and the projected mass is calculated within a radius of $r_\mathrm{2D} = 1 \, \mathrm{kpc}$. The right panel shows the projected density within a range of $r_\mathrm{2D} = 0.25 \, $--$ \, 0.75 \, \mathrm{kpc}$ and the projected mass within $r_\mathrm{2D} = 0.5 \, \mathrm{kpc}$. All subhalos start with the same initial values: in the left panel, $\gamma_\mathrm{2D}\approx-1.2$ and $M_\mathrm{2D}\approx8\times10^{8} \, \mathrm{M}_\odot$, and in the right panel, $\gamma_\mathrm{2D}\approx-0.7$ and $M_\mathrm{2D}\approx3\times10^{8} \, \mathrm{M}_\odot$. The colour indicates the core size at each time. The CDM subhalos always have a green colour, as they do not form a core. The SIDM subhalos produce a much wider range of slopes and masses than the CDM subhalos.

The projected density slope and mass for the CDM subhalos evolve together almost linearly.
Both quantities gradually decrease along a tidal track as tidal forces strip away mass.
In contrast, SIDM subhalos follow a more complex, non-linear evolution. During the core expansion phase, the projected density slope increases. As the core collapses, this trend reverses: the projected density slope decreases and the projected mass increases. This evolution can be seen from the rSIDM subhalo on the outer orbit ({\Large\color{mycolor3} \upperhalfdiamond}). In the left panel, once the core has nearly collapsed, the pattern shifts again: the projected density slope increases, and the projected mass decreases.
This final behaviour is not seen in the right panel indicating that the density profile continues to steepen in the inner region of the halo. Apart from this, the right panel shows a similar evolution for the same quantities in a more central region. 

The other rSIDM subhalos show a stronger decrease in the projected mass as they experience a greater mass loss. This mass loss is strongest when the SIDM subhalo is around its pericentre. During this period, the projected density slope remains constant, whereas the projected mass decreases. Overall, we find steeper projected density slopes and lower masses for closer orbits.

A subhalo with a steep projected density slope and a high projected mass was detected in the gravitational lens SDSSJ0946+1006 \citep{Vegetti_2010} with $M_\mathrm{2D} = (2.2 \, $--$\,3.4)\times10^9 \ \mathrm{M}_\odot$ and $\gamma_\mathrm{2D} < -1.75$ \citep{Minor_2021,2025ApJ...981....2M}. This observation is difficult to explain with a dark substructure alone, but it could be consistent with a faint galaxy \citep{He:2025wco}. In the left panel of Fig.~\ref{fig:ProjectedMass/Slope}, the rSIDM subhalo on the outer orbit ({\Large\color{mycolor3} \upperhalfdiamond}) demonstrates that SIDM could reproduce the observed projected density slope and projected mass. Although the projected mass is somewhat lower than observed, we estimate that a slightly higher initial halo mass, $M_\mathrm{vir} = 3 \times 10^{10} \, \mathrm{M}_\odot$, could bring it into agreement with observations \citep{Li_2025}. Further, observed gravitational lens JVAS B1938+666 \citep{Vegetti_2012} has led to the measurement of two other subhalos, which are modelled well by steep inner density slopes \citep{Despali_2025, Powell_2025, Vegetti_2026}, which CDM struggles to produce. 

In summary, our results show that core collapse in SIDM consistently produces much steeper projected density slopes than in CDM. However, projected mass is highly sensitive to the orbit of the subhalo, the magnitude of the self-interaction cross section, and its velocity dependence. Greater tidal mass loss tends to make the projected density slope steeper but also lowers the projected mass. For the gravitational lens SDSSJ0946+1006, the notable feature is the steepness of the projected density slope rather than the exact value of the projected mass.

\subsubsection{Highly anisotropic scattering}
\label{sec:fSIDM}
We now study the angular dependence of the self-interaction cross section. We examine how forward-dominated self-interactions (fSIDM) modify the subhalo evolution compared to isotropic self-interactions (rSIDM). These two cases represent the extremes for the angular dependence of the cross section, bracketing the range of possible angular dependencies of SIDM and their influence on the subhalo evolution.

We use the same value of the effective cross section for rSIDM and fSIDM, matched through the viscosity cross section. Choosing the same value ensures the same internal core evolution for an isolated halo (see right panel of Fig.~\ref{fig:Cross-section}). Since the effective cross section is matched through the viscosity cross section, the momentum transfer cross section is larger for fSIDM than for rSIDM. The SSHI process scales with the transfer rather than the viscosity cross section. Consequently, fSIDM subhalos are more strongly affected by the SSHI process, typically developing larger cores and experiencing greater delays in core collapse.

\begin{figure*}
    \centering
    \includegraphics[width=\columnwidth]{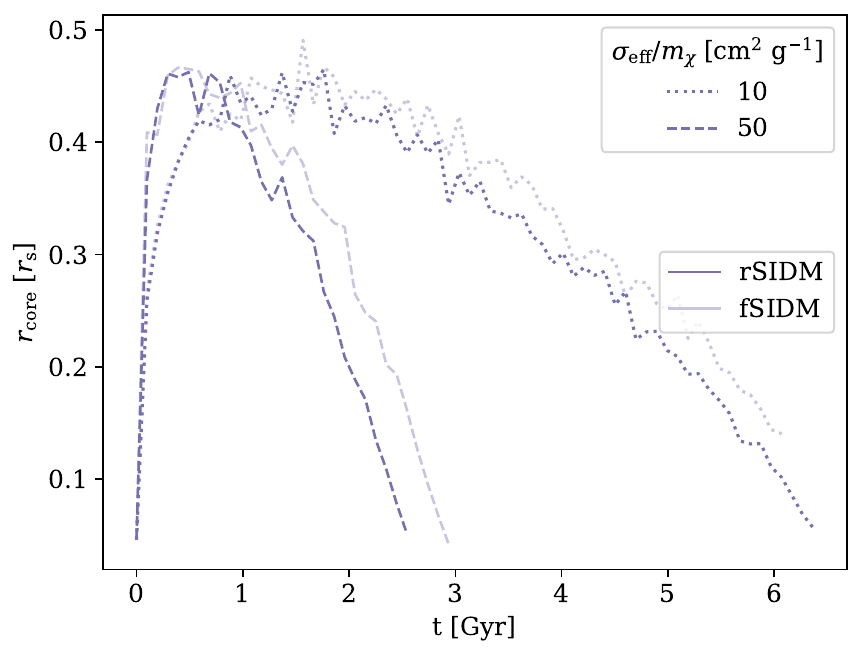}
    \includegraphics[width=\columnwidth]{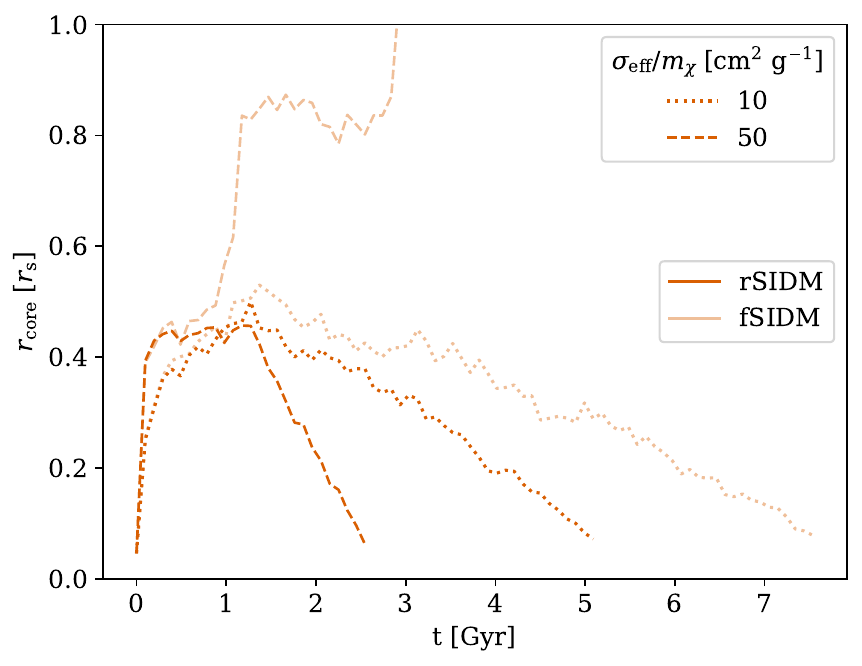}
    \includegraphics[width=\columnwidth]{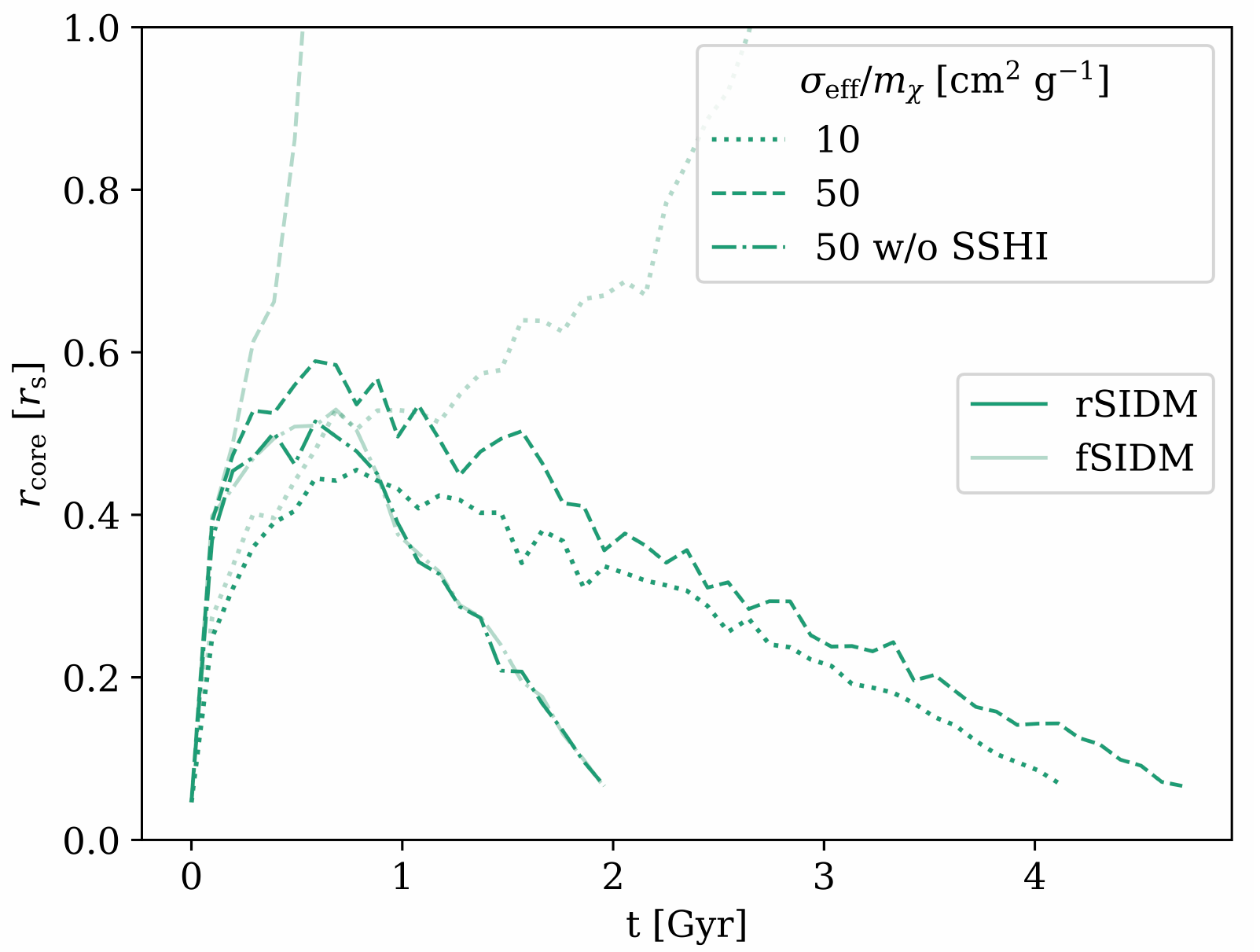}
    \includegraphics[width=\columnwidth]{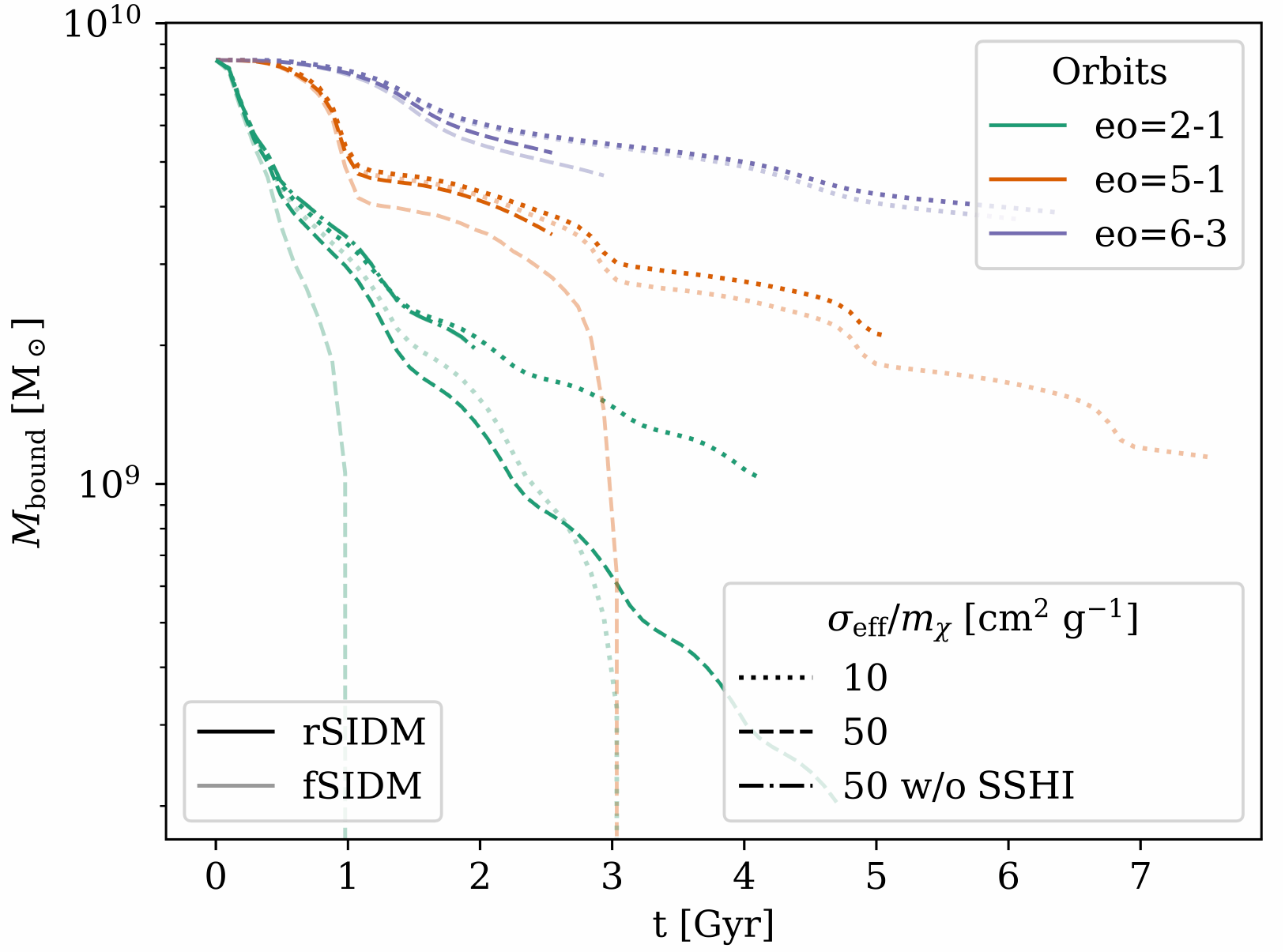}
    \caption{The evolution of the core size $r_\mathrm{core}$ and the bound mass $M_\mathrm{bound}$ (bottom right panel). The core size evolution is shown for three orbits (associated with the colour in the legend in the bottom right panel). For each orbit, we show the simulations with isotropic scattering (rSIDM) and far-forward scattering (fSIDM), each with two different effective cross sections. The bottom left panel shows the core evolution of the inner orbits and includes the evolution of the rSIDM subhalo without the SSHI process and with the higher effective cross section. The radius $r_\mathrm{core}$ is given in units of the scale radius of the initial NFW subhalo.}
    \label{fig:CoreEvolution_fSIDM}
\end{figure*}

Figure~\ref{fig:CoreEvolution_fSIDM} shows the evolution of the core size for the subhalos on three different orbits. Each orbit is displayed in a separate panel. The decay of the orbit is slightly larger for fSIDM. The rSIDM subhalos are the bolder dotted or dashed lines, and fSIDM the fainter dotted or dashed lines. In all cases, the fSIDM subhalos reach larger or equal core sizes than their rSIDM counterparts, with larger differences for orbits that pass closer to the host halo. Since the SSHI process and tidal heating are strongest near the pericentre, the differences between fSIDM and rSIDM subhalos are most pronounced during pericentric passages.

For fSIDM subhalos on the inner orbit, the core expansion phase does not end. As the core grows, the subhalo becomes progressively less gravitationally bound, leading to its disruption by the tidal forces of the host system.
Consequently, as shown in the bottom right of Fig.~\ref{fig:CoreEvolution_fSIDM}, fSIDM subhalos lose more mass over time than rSIDM subhalos. This increased mass loss, primarily in the outskirts, increases the outward heat flow and therefore accelerates the core collapse. However, the increased effect of the SSHI process for fSIDM delays the core collapse more strongly. Additional results for fSIDM are presented in Appendix~\ref{sec:additional_fSIDM}.
Our comparisons highlight the extreme sensitivity of fSIDM subhalos to environmental effects, far beyond what is seen in the rSIDM scenario. As a consequence, satellites are an interesting probe of the angular dependence of the self-interaction cross section.

\section{Discussion}
\label{sec:discussion}
In this section, we discuss the current limitations and future improvements of the simulation setup to model the SSHI process of the SIDM subhalos, particularly how the interactions between the subhalo and the host halo particles are handled. Additionally, we reflect on the assumptions underlying the Eddington inversion method and suggest directions for more realistic modelling.

When simulating the SSHI process, various outcomes can occur when DM subhalo and DM host halo particles scatter. These scenarios are:
\begin{enumerate}
    \item Both particles are unbound to the subhalo after scattering.

    \item Both particles are bound to the subhalo after scattering.

    \item The subhalo particle remains bound, while the host halo particle remains unbound.

    \item The subhalo particle becomes unbound, while the host halo particle becomes bound.
\end{enumerate}
Scenarios~3 and~4 are physically indistinguishable for elastic scattering, because the DM particles are fundamentally identical. The only difference lies in the labelling used in the simulation.
We account for scenarios~1 and~3 in our current simulation setup. In both cases, the subhalo particle continues to be simulated, while the virtual host halo particle is discarded.

Scenario~2 is not implemented for this work. To capture this case correctly, the simulation must convert the virtual particle into a subhalo particle after the interaction. The probability that both the particles are bound after scattering is much smaller than one of them being bound, and hence including of Scenario~2 won't change our results. 

\textsc{OpenGadget3} itself includes an implementation for scenario 4. If the scattering angle exceeds $90^\circ$, the host halo particle is more likely to become bound, while the subhalo particle becomes unbound. In this case, the simulation maps the scattering angle $\theta$ when larger than $\uppi/2$ to a scattering angle of $\uppi - \theta$ \citep[mimicing switching particle labels, see appendix~A by][]{Fischer_2021b}.
This procedure is only an approximation and does not capture the full physical details of the interaction.

We have implemented the Eddington inversion under the assumptions of spherical symmetry and an isotropic velocity dispersion. Given that host systems likely possesses a stellar disks or ellipsoids and exhibit a non-isotropic velocity distribution, generalising the inversion formalism to accommodate axial symmetry and arbitrary velocity dispersion profiles would be appropriate to more accurately capture the dynamical structure of subhalos within realistic galactic host systems. In addition, a time-dependent potential would make it possible to capture mass growth of the host system or gravitational interactions with other galaxies. This would be useful for extensions of our work to the Milky Way and Andromeda, and enable comparison with observations of satellites. Stars in these satellites respond to changes to the gravitational potential due to the effects we have described, and the kinematics and density profiles of the stars in faint satellites galaxies could provide a way to test the predictions for SIDM models.

As we neglected dynamical friction, our simulation set-up is restricted to systems that fulfil $m_\mathrm{sub} \leq 1/1000 \ m_\mathrm{host}$. Within this regime, the framework can be used to explore a range of host–subhalo mass ratios. This can be extended by implementing dynamical friction. The drag force influences the subhalo trajectories based on their mass and orbital characteristics. The deceleration due to this force can be calculated as in~\citet{Slone_2023} using the Chandrasekhar formula. The calculation requires identifying bound particles during the simulation run and computing the properties of the satellite, such as its mass $m_\mathrm{sat}$ and velocity $\vec{v}_\mathrm{sat}$.

Our results on the evolution of SIDM subhalos have direct implications for analytic modelling, such as the parametric model of \citet{Yang_2024}. For SIDM subhalos experiencing weak tidal forces, we find that during core expansion, the density slope at radii a little smaller than the scale radius becomes steeper while the mass enclosed within those radii remains roughly constant (see left panel of Fig.~\ref{fig:ProjectedMass/Slope} for the projected quantities at $2/3 \, r_\mathrm{s}$). At even smaller radii, the density slope becomes shallower, and the mass decreases (see right panel of Fig.~\ref{fig:ProjectedMass/Slope} for the projected quantities at $1/3 \, r_\mathrm{s}$). During core collapse, the slope of the density steepens over a substantial radial range, with exceptions at the very center of the halo and large radii ($r \gtrsim r_\mathrm{s}$). The steepening of the slope is accompanied by an increase in the mass enclosed within these intermediate radii. This behaviour continues up to a turnaround point in time, beyond which the slope begins to flatten and the mass enclosed decreases. This turnaround point is reached earlier for larger radii \citep[see also][]{Fischer_2025}.
For subhalos experiencing strong tidal forces, the mass can decrease even before the turnaround point while the inner slope steepens further. These results highlight an important modelling principle: during core collapse, analytic models should allow the inner slope within $r \lesssim r_\mathrm{s}$, except at the center, to steepen further, and permit reductions in the inner mass when physically justified by strong tidal forces (see Appendix~\ref{sec:additional_fSIDM}).

\section{Conclusion}
\label{sec:conclusion}
In this work, we explore the evolution of SIDM subhalos, focusing on three key environmental processes: the SSHI process, tidal stripping, and tidal heating. We have developed and implemented a dedicated method for modelling the SSHI process using virtual host halo particles. This setup allowed us to efficiently run high-resolution simulations of SIDM subhalos in realistic host environments.

Our simulations show that SIDM subhalos follow a more diverse and non-linear evolutionary path than their CDM counterparts. This complexity arises from the interplay between self-interactions and environmental effects. 

Our main findings are as follows:
\begin{enumerate}
    \item The SSHI process from host–subhalo scattering can extend the core expansion phase or delay core collapse, in agreement with \citet{Zeng_2022}.

    \item Tidal heating can lead to a renewed and short-lived core expansion phase or delay core collapse, consistent with \citet{Zeng_2022}.

    \item Although the SSHI process and tidal heating can slow down the core evolution, tidal stripping accelerates the core evolution by steepening the velocity dispersion at the outskirts and thereby increasing the heat outflow. This is in agreement with \citet{Nishikawa_2020}.

    \item The SSHI process, tidal stripping, and tidal heating influence the structural evolution of the subhalo more strongly for SIDM models with far-forward scattering than with isotropic scattering. 

    \item Tidal stripping is enhanced in SIDM due to the formation of low-density cores. However, even for the large self-interaction cross section we simulated, most subhalos survive.

    \item All the SIDM subhalos we simulated achieve steep central profiles at late times, which has exciting implications for gravitational lensing by subhalos.

\end{enumerate}

Simulations with analytic host halos and virtual particles to capture the SSHI process provide high-fidelity modelling of low-mass SIDM subhalos and enable efficient exploration of the SIDM model parameter space.
Our results highlight the physical richness of SIDM models and the expanded range of values predicted for the structural parameters of subhalos. Our implementation demonstrates a viable path forward for simulating SIDM in realistic environments and constraining its properties through detailed comparisons of structural properties of subhalos and satellites galaxies with observational data.

\begin{acknowledgements}
We thank all participants of the Darkium SIDM Journal Club for the discussion.
MSF is delighted to thank Klaus Dolag for helpful discussions on \textsc{SUBFIND} and other technical aspects.
MSF acknowledges support by the \emph{Deutsche Forschungsgemeinschaft (DFG, German Research Foundation)} under Germany’s Excellence Strategy -- EXC-2094 ``Origins'' -- 390783311 and the COMPLEX project from the European Research Council (ERC) under the European Union’s Horizon 2020 research and innovation program grant agreement ERC-2019-AdG 882679 and gratefully acknowledges the support of the Alexander von Humboldt Foundation through a Feodor Lynen Research Fellowship.
KB acknowledges support from the National Science Foundation under Grant No.~PHY-2413016. MK acknowledges support from the National Science Foundation under Grant No.~PHY-2210283 and PHY-2514888.

Software:
NumPy \citep{NumPy},
Matplotlib \citep{Matplotlib}.
\end{acknowledgements}

\bibliographystyle{aa}
\bibliography{bib.bib}

\begin{appendix}
\section{The SSHI process}
\label{sec:DensityMassPofiles}
Here, we examine the evolution and structure of the DM subhalos to understand the impact of self-interactions, in particular on the SSHI process.
In the following analysis, we calculate for each time step the current scale density $\rho_\mathrm{s}(t)$ and scale radius $r_\mathrm{s}(t)$ from the CDM simulations with the relation
\begin{equation}
    \rho_\mathrm{s}(t)=\frac{1}{4 \, \mathrm{G}}\left( \frac{v_\mathrm{max}(t)}{1.64835 \, r_\mathrm{s}} \right)^2
    \label{eq:rho_s}
\end{equation}
and $r_\mathrm{s}(t)=r_\mathrm{max}(t)/2.1626$ \citep{Yang_2024}. These scale density and radius are then used to normalise the units in both the CDM and the corresponding SIDM simulations.

\begin{figure*}
    \centering
    \includegraphics[width=0.85\textwidth]{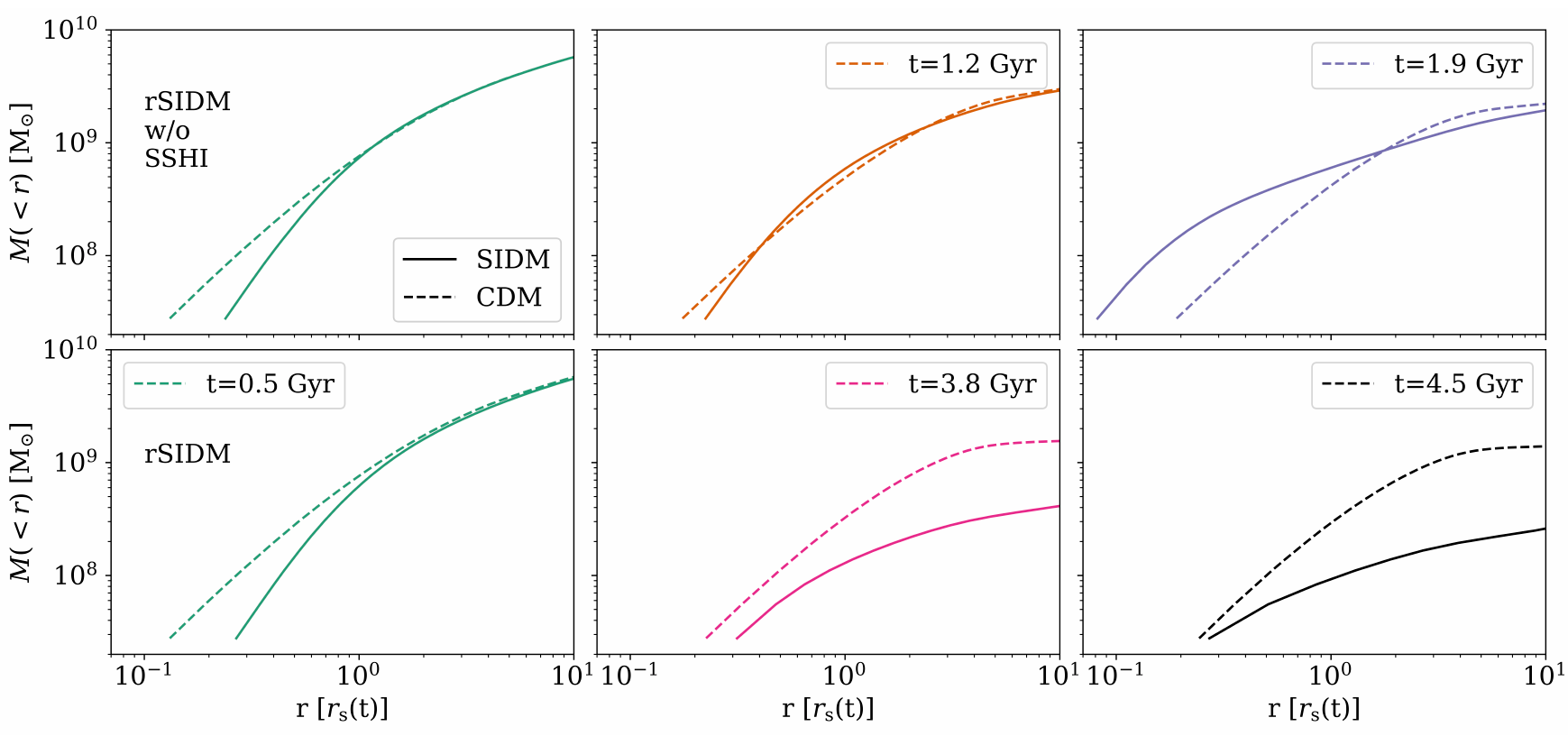}
    \caption{Mass profile of the isotropic SIDM subhalos (solid lines) with the higher effective cross section ($\sigma_\mathrm{eff}/m_\chi = 50 \  \mathrm{cm}^2 \ \mathrm{g}^{-1}$) on the inner orbit (eo=2--1): top panel without the SSHI process, bottom panel with the SSHI process. The dashed lines show the mass profile of the corresponding CDM subhalo. The radius $r$ is given in units of the scale radius, which is calculated at each time step of the corresponding CDM subhalo (see Eq.~\eqref{eq:rho_s}).}
    \label{fig:MassProfile}
\end{figure*}

First, we focus on the DM subhalo on the inner orbit. We compare CDM and rSIDM with and without the SSHI process for the higher cross section. Figure~\ref{fig:MassProfile} shows the evolution of the enclosed mass as a function of radius. The radius is normalised by the scale radius at each time step from the corresponding CDM simulation. As the SIDM subhalo without the SSHI process (upper panels) evolves a core, its mass profile becomes reduced in the central region compared to the CDM simulation (dashed). In the outer regions, both profiles remain similar. During the core collapse phase, the overall mass profile appears similar, although subtle structural differences persist. After the core collapse, the SIDM subhalo contains significantly more mass concentrated in the central region.

When the SSHI process is included (lower panels), the overall mass profile progressively reduces over time. Even after the core collapse, the central mass remains lower than in the CDM simulation due to the continued mass loss from the SSHI process and the enhanced tidal mass loss from the extended core evolution.

\begin{figure*}
    \centering
    \includegraphics[width=0.85\textwidth]{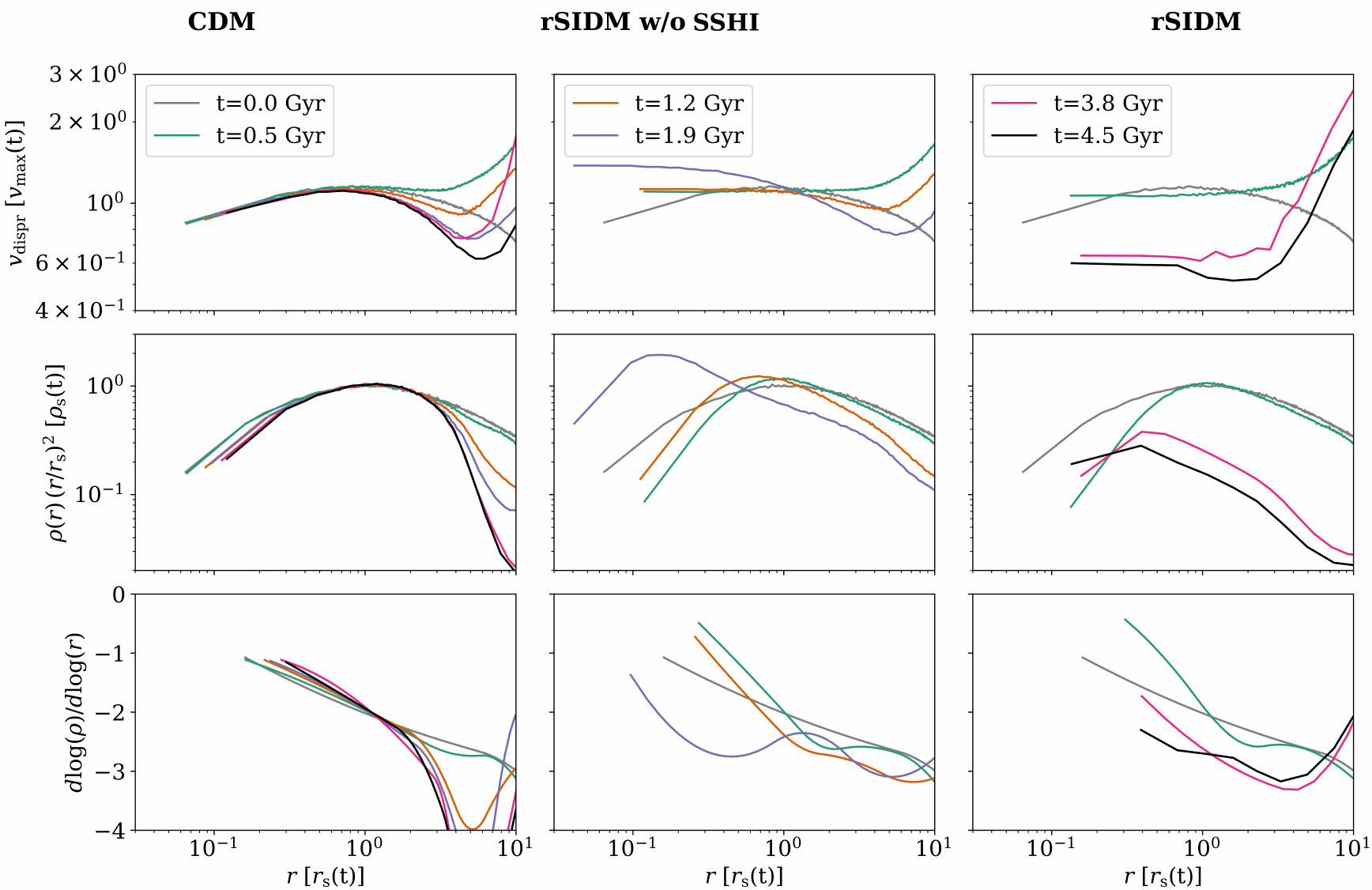}
    \caption{All DM subhalos shown here are from simulations of the inner orbit (eo=2--1) with the higher isotropic cross section ($\sigma_\mathrm{eff}/m_\chi = 50 \  \mathrm{cm}^2 \ \mathrm{g}^{-1}$). The left column corresponds to the CDM simulation. The middle column shows the rSIDM simulation without the SSHI process. The right column illustrates the rSIDM simulation, including the SSHI process. The upper row shows the velocity dispersion as a function of radius. The middle row displays the density profile times the radius squared of the subhalo. The bottom row presents the logarithmic slope of the density profile with respect to the radius. The radius $r$ is given in units of the scale radius, which is calculated at each time step of the corresponding CDM subhalo as $v_\mathrm{max}$ and $\rho_\mathrm{s}$ (see Eq.~\eqref{eq:rho_s}).}
    \label{fig:TimeEvolution}
\end{figure*}

In Fig.~\ref{fig:TimeEvolution}, we compare the evolution of the velocity dispersion profile (top row), the density profile times radius squared (middle row), and the logarithmic density slope profile (bottom row). The density profile is given in terms of the scale density $\rho_\mathrm{s}$, which is recalculated from the CDM subhalo at each time step (see Eq.~\eqref{eq:rho_s}).
CDM subhalos (left column) exhibit small changes in the central regions of their profiles, while tidal stripping influences them in the outer regions by steadily reducing the velocity and density. As a result, the logarithmic density slope profile stays nearly constant, with a noticeable decrease only in the outskirts.
In the absence of the SSHI process (middle column), the SIDM subhalo evolves similarly to an isolated halo; aside from mass loss in the outskirts, its central structure changes quite similarly. The density slope profile becomes shallower in the central regions due to core formation, while the outer slope becomes steeper. As the core collapses, the region of shallow slope shifts inward.
In contrast, when the SSHI process is included (right column), the SIDM subhalo undergoes substantial structural changes, especially in the central regions, where the density and velocity dispersion are notably reduced. The inclusion of the SSHI process leads to an overall steeper density slope across the profile. Those significant differences in the SIDM subhalo evolution highlight the SSHI process as a key mechanism driving the deviation of SIDM subhalos from their isolated counterparts.

\begin{figure*}
    \centering
    \includegraphics[width=0.9\columnwidth]{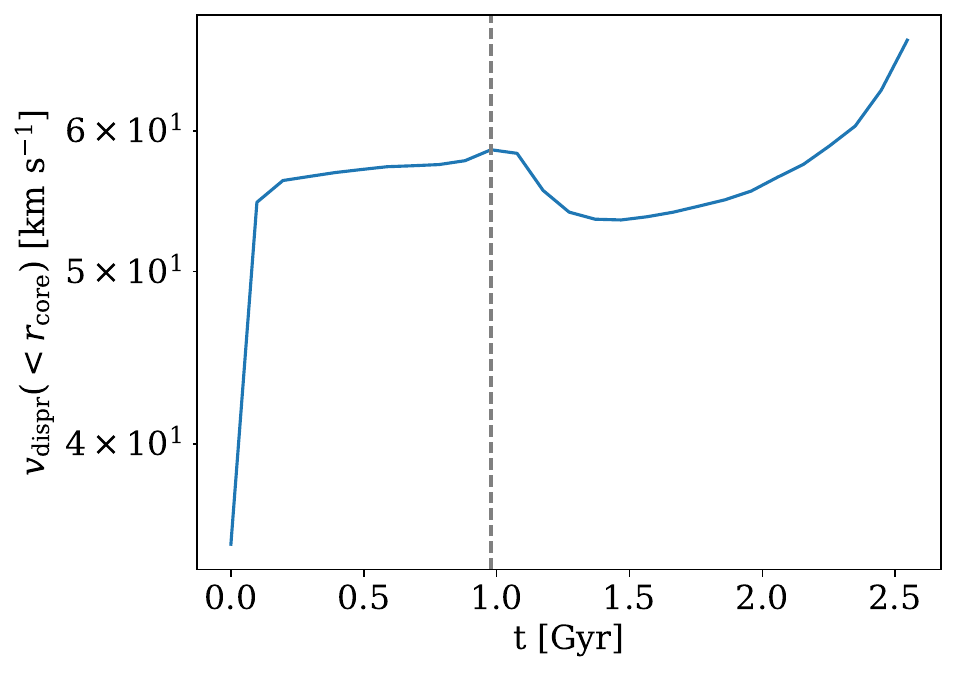}
    \includegraphics[width=1.1\columnwidth]{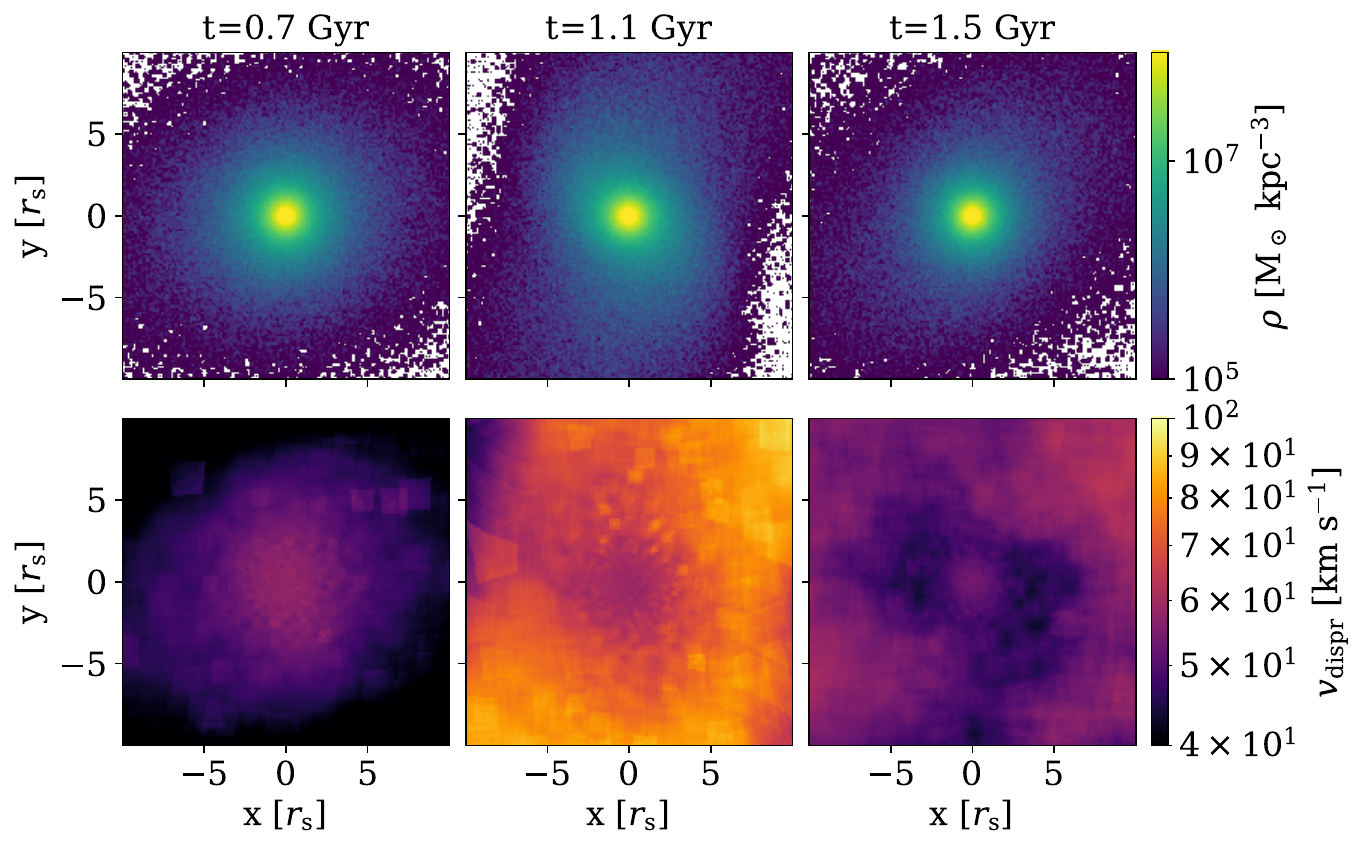}
    \caption{The DM subhalo shown here is from the simulation on the intermediate orbit (eo=5--1) with the higher isotropic cross section ($\sigma_\mathrm{eff}/m_\chi = 50 \ \mathrm{cm}^2 \ \mathrm{g}^{-1}$). The left panel shows the velocity dispersion within the core. The grey dashed line indicates the pericentre. The right panel shows, in the upper row, the density map and, in the lower row, the velocity dispersion map. The coordinates $x$ and $y$ are given in units of the scale radius of the initial NFW subhalo.}
    \label{fig:subhalo_velocity_dispersion_map}
\end{figure*}

Figure~\ref{fig:subhalo_velocity_dispersion_map} illustrates the effect of tidal heating on the velocity dispersion of a SIDM subhalo (complementing the main discussion in Sect.~\ref{sec:TidalHeating}). The left panel shows the evolution of the velocity dispersion within the core, while the right panel presents the density and velocity dispersion maps around the pericentre. During the pericentre ($t=1.0 \, \mathrm{Gyr}$), the core velocity dispersion increases slightly, followed by a sharp decline. The velocity dispersion map at $t=1.1 \, \mathrm{Gyr}$ shows that tidal forces increase the velocity dispersion throughout the subhalo, which increases toward the outskirts. After pericentre ($t=1.5 \, \mathrm{Gyr}$), the overall velocity dispersion decreases compared to the pre-pericentre state at $t=0.7 \, \mathrm{Gyr}$. Higher velocity dispersions are still seen at the outskirts, caused by stripped particles moving away from the subhalo. While the velocity dispersion map shows significant changes during the pericentre passage, the density map is only weakly affected.

\section{Additional results for highly anisotropic scattering}
\label{sec:additional_fSIDM}

Here we present additional results, focussing on the forward-dominated SIDM model (fSIDM), complementing the main discussion in Sect.~\ref{sec:fSIDM}. Here we also present the key quantities, such as the maximum circular velocity and its radius, and the projected density slope and mass, for the fSIDM subhalo.

The fSIDM subhalos with the higher cross section exhibit slightly stronger orbital decay compared to their rSIDM counterparts. This arises as the SSHI process affects the fSIDM subhalos more strongly. The difference becomes more pronounced for closer orbits, where the SSHI process has a greater impact.

\begin{figure*}
    \centering
    \includegraphics[width=0.85\textwidth]{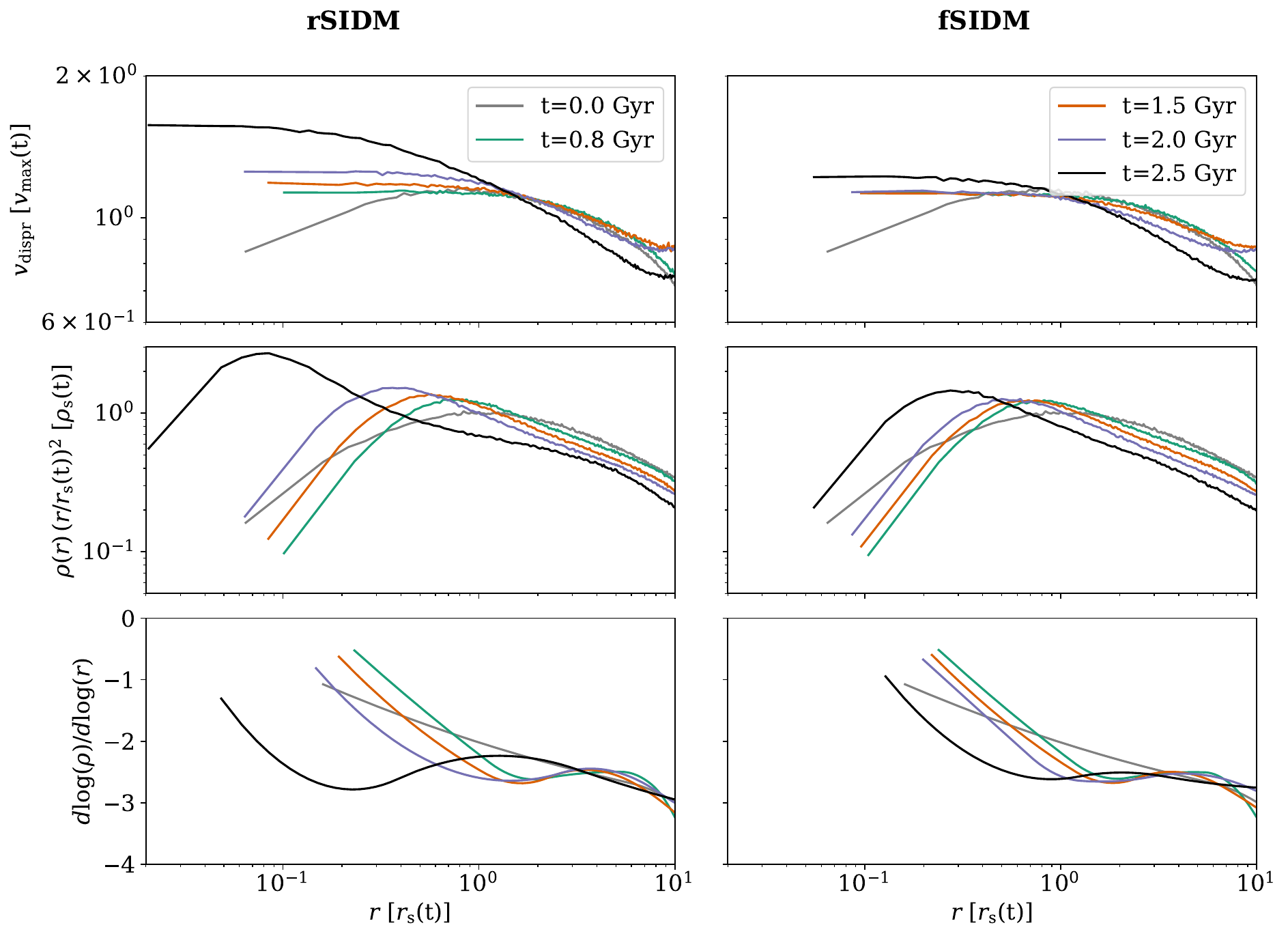}
    \caption{All DM subhalos shown here are from simulations of the outer orbit (eo=6--3) with the higher isotropic cross section ($\sigma_\mathrm{eff}/m_\chi = 50 \ \mathrm{cm}^2 \ \mathrm{g}^{-1}$). The left column corresponds to the rSIDM (isotropic) simulation. The right column shows the fSIDM (forward-dominated) simulation. The upper row shows the velocity dispersion as a function of radius. The middle row displays the density profile of the subhalo. The bottom row presents the logarithmic slope of the density profile with respect to radius. The radius $r$ is given in units of the scale radius, which is calculated at each time step of the corresponding CDM subhalo as $v_\mathrm{max}$ and $\rho_\mathrm{s}$ (see Eq.~\eqref{eq:rho_s}).}
    \label{fig:DensityVelDisprProfileEvolution}
\end{figure*}

\begin{figure*}
    \centering
    \includegraphics[width=1.02\columnwidth]{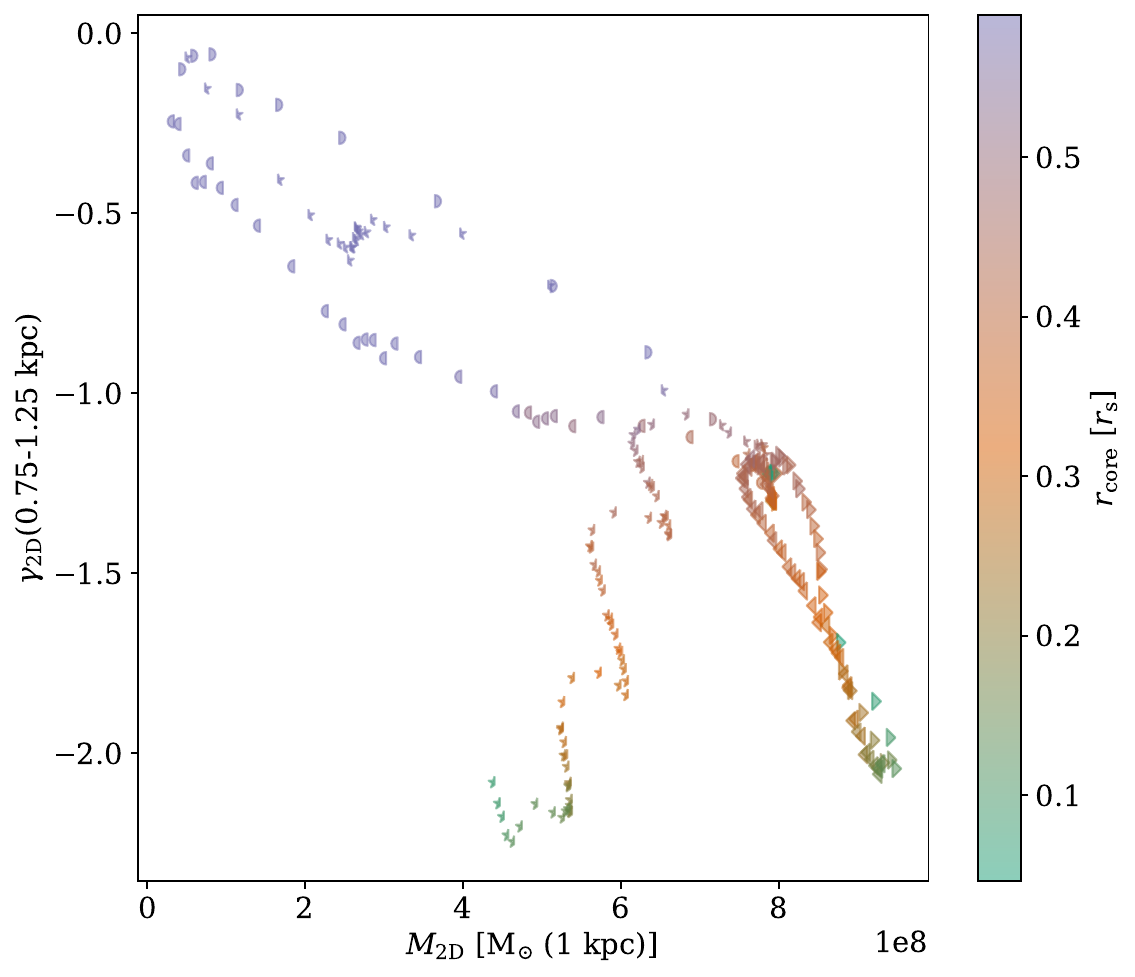}
    \includegraphics[width=0.98\columnwidth]{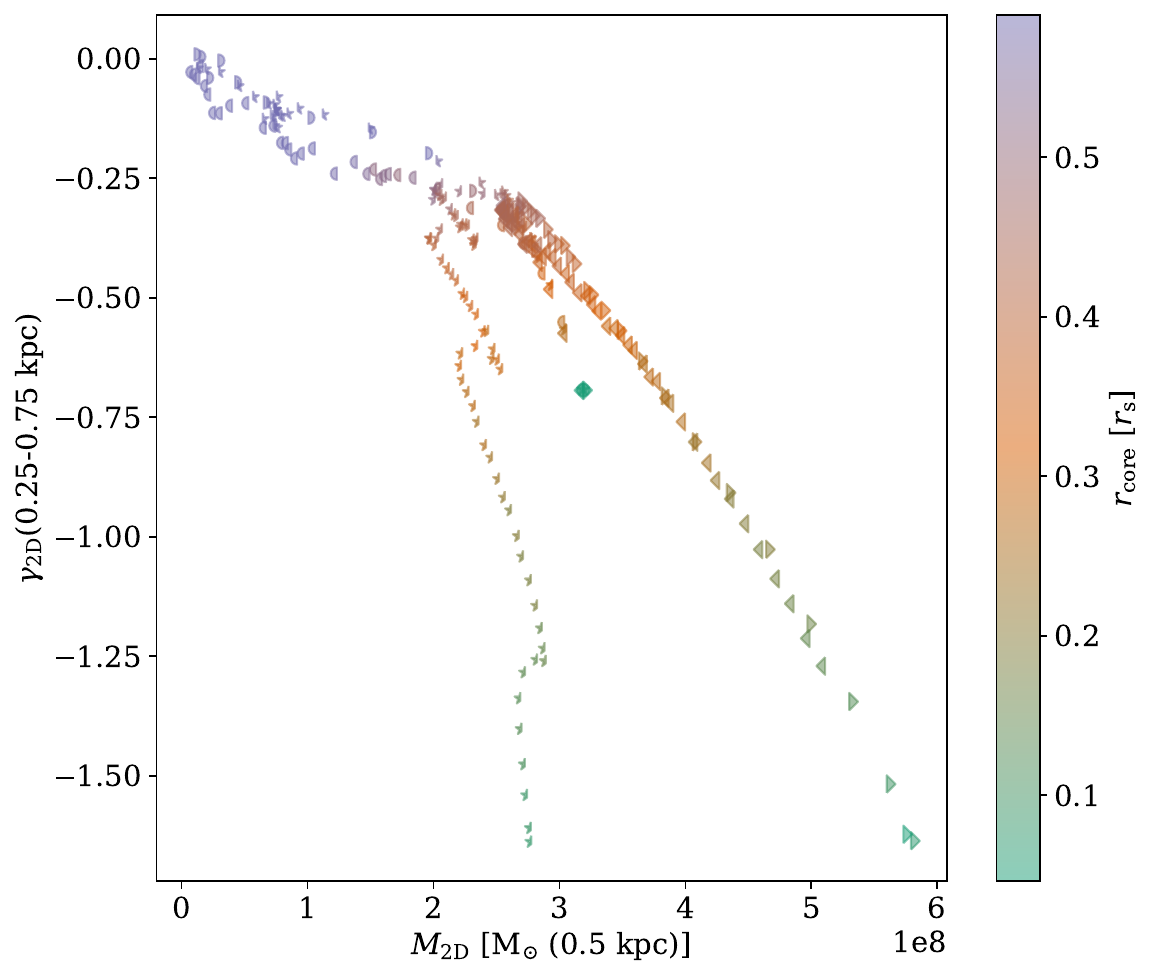}
    \caption{Projected properties of the fSIDM simulations. Left: Projected density slope $\gamma_\mathrm{2D}$ (0.75--1.25 kpc) vs.\ projected mass $M_\mathrm{2D}$ (at 1 kpc). Right: Projected density slope (0.25--0.75 kpc) vs.\ projected mass (at 0.5 kpc). Symbols follow Table~\ref{tab:orbits}. Colour indicates $r_\mathrm{core}$ in units of the initial scale radius.}
    \label{fig:ProjectedMass/Slope_fSIDM}
\end{figure*}

\begin{figure}
    \centering
    \includegraphics[width=\columnwidth]{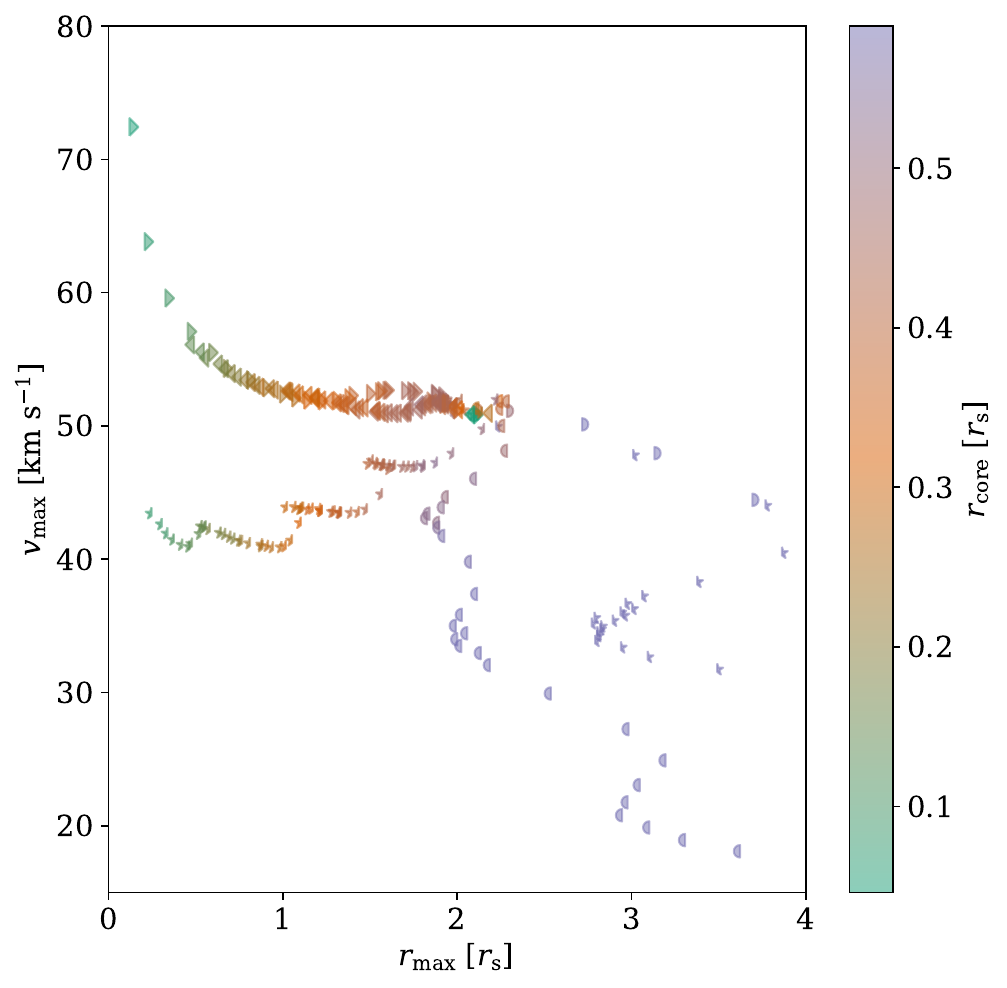}
    \caption{Maximum circular velocity in the fSIDM simulation. Time evolution of maximum circular velocity $v_\mathrm{max}$ and corresponding radius $r_\mathrm{max}$. Symbols follow Table~\ref{tab:orbits}. Colour indicates the DM core size $r_\mathrm{core}$. Radii $r_\mathrm{max}$ and $r_\mathrm{core}$ are in units of the initial NFW scale radius.}
    \label{fig:VMAX_fSIDM}
\end{figure}

Fig.~\ref{fig:DensityVelDisprProfileEvolution} shows the evolution of the SIDM subhalos with the higher cross section on the outer orbit. Those SIDM subhalos have a very similar evolution to the corresponding isolated SIDM halo.
The left column of Fig.~\ref{fig:DensityVelDisprProfileEvolution} shows how the velocity dispersion (top row), density (middle row), and density slope profile (bottom row) evolve for the isotropic cross section. This subhalo shows the decrease in the projected density slope and an increase in the projected mass, which CDM struggles to produce. 
The right column of Fig.~\ref{fig:DensityVelDisprProfileEvolution} shows the corresponding fSIDM subhalo (forward-dominated cross section), which has a slightly lower projected mass. For $t=0.8 \ \mathrm{Gyr}$, both SIDM subhalos have evolved a core and now go into core collapse. For $t=1.5 \ \mathrm{Gyr}$, we can see how the core collapse proceeds by the increasing central density. For the rSIDM subhalo, we see, as expected, that the central velocity dispersion increases. However, for fSIDM, we observe a slight decrease in the central velocity dispersion, which is linked to its enhanced mass loss compared to rSIDM. As a result, the maximum circular velocity of the fSIDM subhalo is slightly lower (Fig.~\ref{fig:VMAX_fSIDM}), as is its projected mass (Fig.~\ref{fig:ProjectedMass/Slope_fSIDM}). For $t=2.5 \ \mathrm{Gyr}$, both SIDM subhalos show a strong increase in central density and central velocity dispersion, which shows that both subhalos are in the late stage of core collapse.

The bottom row shows the evolution of the density slope profile and helps to understand how internal and external processes shape the density slope profile. The initial NFW profile is almost linear (grey line). For the rSIDM subhalo, as the core evolves, the profile becomes S-shaped: it first decreases to a minimum, then rises to a maximum, and finally decreases again. During core collapse, the S-shape moves inward, the minimum value decreases, and the maximum value increases. For the fSIDM subhalo, where external effects play a stronger role, the S-shape shifts less inwards, and the minimum and maximum values remain roughly constant.

\begin{figure*}
    \centering
    \includegraphics[width=\columnwidth]{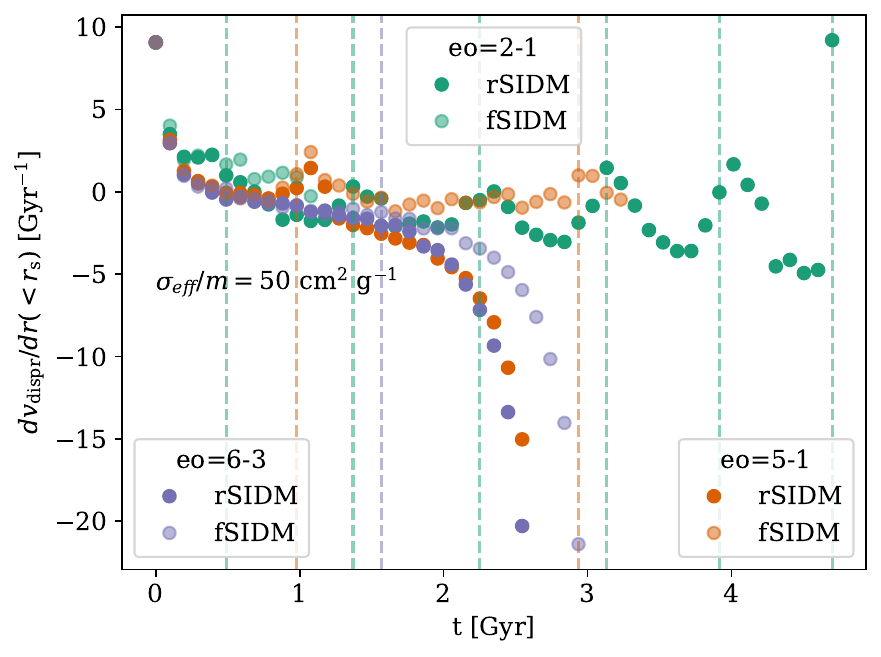}
    \includegraphics[width=\columnwidth]{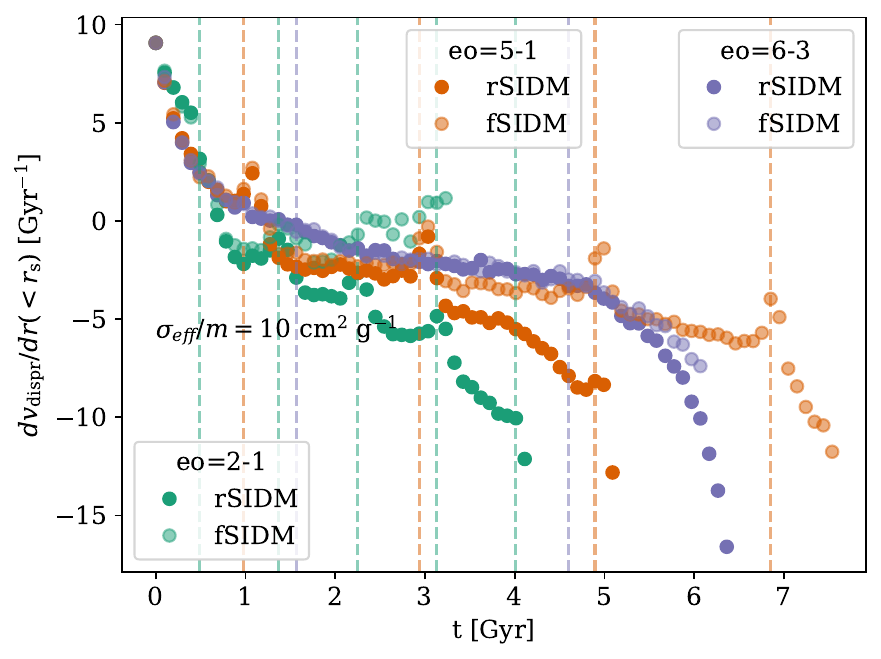}
    \caption{Gradient of velocity dispersion within the initial $r_\mathrm{s}$ of the subhalos with: 
    left: $\sigma_\mathrm{eff}/m_\chi = 50 \ \mathrm{cm}^2 \ \mathrm{g}^{-1}$,
    right: $\sigma_\mathrm{eff}/m_\chi = 10 \ \mathrm{cm}^2 \ \mathrm{g}^{-1}$.
    Dashed vertical lines indicate the pericentre passage.}
    \label{fig:HeatFlow_fSIDM}
\end{figure*}

Figure~\ref{fig:HeatFlow_fSIDM} shows the evolution of the gradient of the velocity dispersion, which indicates if the subhalo is in core expansion phase (for positive values) or in core collapse phase (for negative values). The left panel shows subhalos with the higher cross section ($\sigma_\mathrm{eff}/m_\chi = 50 \ \mathrm{cm}^2 \ \mathrm{g}^{-1}$), and the right panel with the lower cross section ($\sigma_\mathrm{eff}/m_\chi = 10 \ \mathrm{cm}^2 \ \mathrm{g}^{-1}$). The fainter points represent rSIDM subhalos, while bolder points indicate fSIDM. Coloured vertical lines mark the pericentres of each orbit. Around pericentre passages, fSIDM subhalos consistently exhibit steeper gradients than corresponding rSIDM subhalos. These differences are more pronounced for tighter orbits, highlighting the stronger influence of tidal heating and the SSHI process on fSIDM.

\begin{figure*}
    \centering
    \includegraphics[width=\columnwidth]{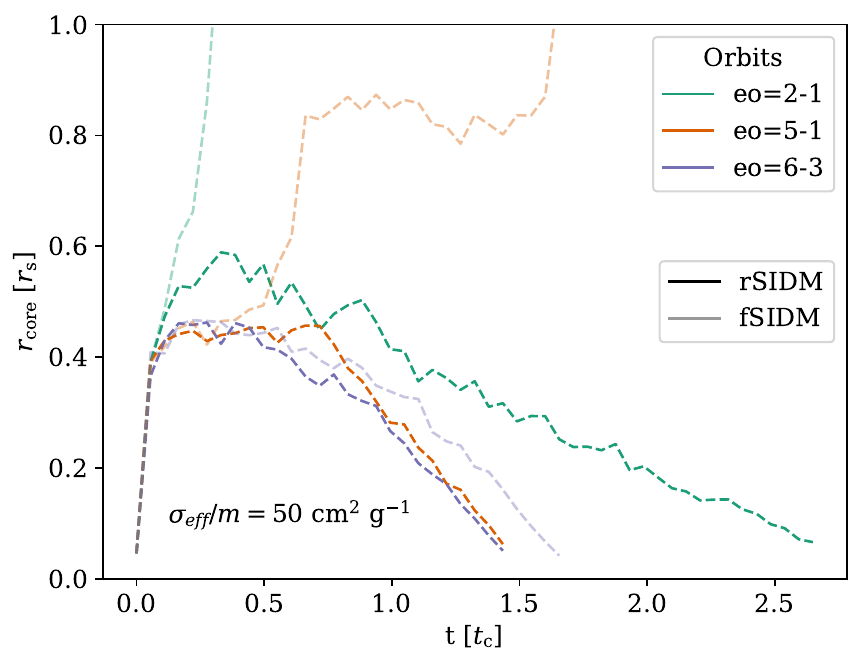}
    \includegraphics[width=\columnwidth]{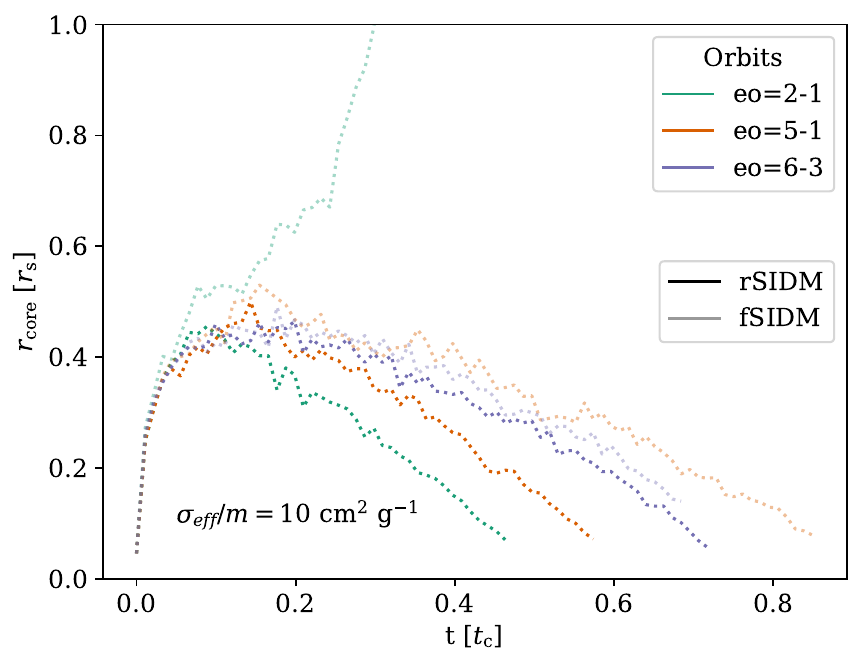}
    \caption{Core size as a function of time, normalized by the collapse time $t_\mathrm{c}$. The radius $r_\mathrm{core}$ is given in units of the scale radius of the initial NFW subhalo. Left panel: High cross section simulations. Right panel: Low cross section simulations.}
    \label{fig:NormCoreEvolution}
\end{figure*}

Figure~\ref{fig:NormCoreEvolution} shows the core evolution as a function of time, normalized by the collapse time \citep{Yang_2024}
\begin{equation}
    t_\mathrm{c}=\frac{150}{C}\frac{1}{(\sigma_\mathrm{eff}/m_\chi) \, (4 \, \rho_\mathrm{s}) \, r_\mathrm{s}}\frac{1}{\sqrt{4 \, \uppi \, \mathrm{G} \, (4 \, \rho_\mathrm{s})}} \,,
\end{equation}
where the calibration constant is set to $C=0.75$.
For the higher isotropic cross section, we can see that the two simulations on the outer orbits collapse almost as an isolated halo. For the simulation on the inner orbit, we see a strongly extended core evolution. 
For the lower isotropic cross section, we see a shortened core evolution. The closer the orbit is to the host, the shorter the core evolution is. 
The forward-dominated cross section simulations exhibit an extended core evolution compared to their isotropic counterparts. This extended core evolution increases for closer orbits.

The impact of fSIDM is also visible in the projected mass and density slope. Figure~\ref{fig:ProjectedMass/Slope_fSIDM} shows the projected logarithmic density slope $\gamma_\mathrm{2D}$ versus projected mass $M_\mathrm{2D}$ (left panel) and the time evolution of $\gamma_\mathrm{2D}$ (right panel). These quantities are calculated from a face-on view, perpendicular to the orbital plane. The colour encodes the DM core size $r_\mathrm{core}$. The fSIDM subhalos, due to their sensitivity to environmental effects (tidal stripping and the SSHI process), have lower projected masses than their corresponding rSIDM subhalos (see Fig.~\ref{fig:ProjectedMass/Slope}), while their projected density slope is roughly the same.

Figure~\ref{fig:VMAX_fSIDM} shows the evolution of the maximum circular velocity $v_\mathrm{max}$ and the radius $r_\mathrm{max}$ at which it occurs. The fSIDM subhalos span a wider range of maximum circular velocities, including values smaller than those found in rSIDM. The corresponding radii also cover a wider range, with especially larger values compared to rSIDM (see Fig.~\ref{fig:VMAX}). The fSIDM subhalos have lower maximum circular velocity than their corresponding rSIDM subhalo, due to the enhanced mass loss by tidal stripping and the SSHI process.

The wide range of key quantities for fSIDM reflects its stronger sensitivity to tidal and SSHI process effects. This arises because the cross section is matched using the viscosity cross section, ensuring the same internal core evolution, while the SSHI process scales with the transfer cross section, which is stronger for fSIDM than for rSIDM. Since rSIDM and fSIDM represent the two limiting cases of angular dependence, the key properties of the corresponding subhalos span the full range of expected outcomes. In this sense, the values for rSIDM and fSIDM bracket the possible range of each quantity.

\begin{figure*}
    \centering
    \includegraphics[width=\columnwidth]{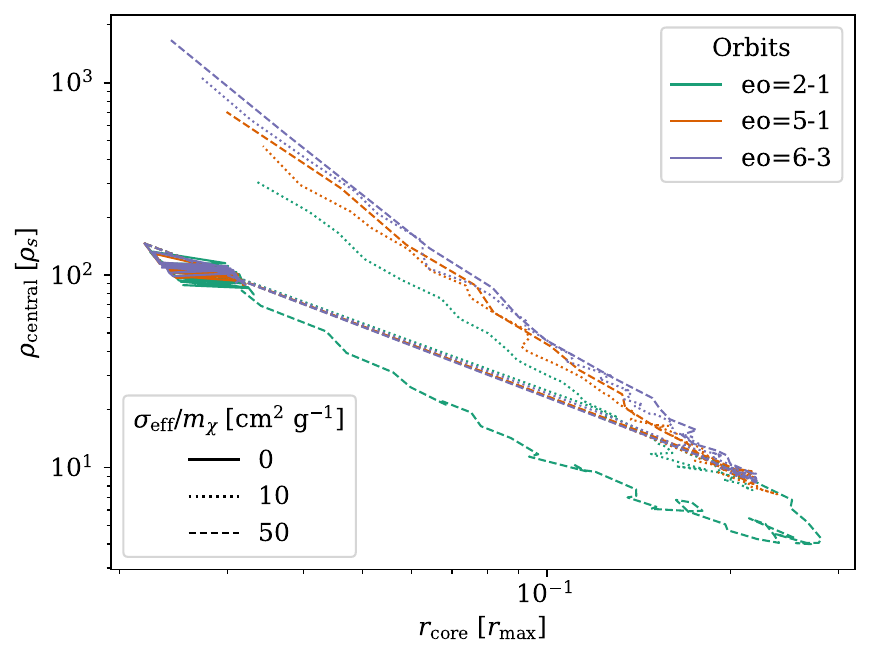}
    \includegraphics[width=\columnwidth]{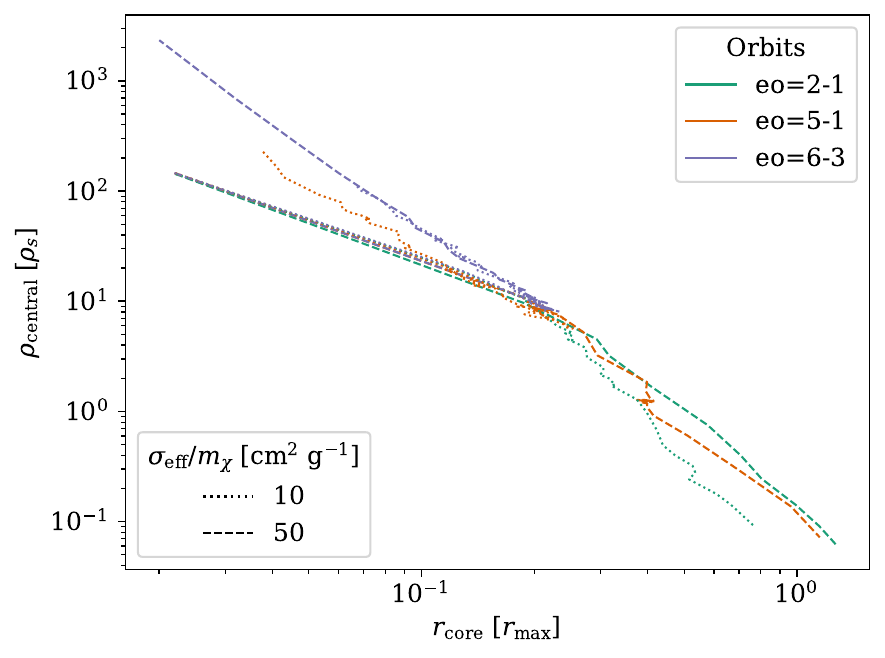}
    \caption{Central density vs.\ core size: left: CDM and isotropic SIDM subhalos, right: forward-dominated SIDM subhalos. Central density is given in terms of the initial $\rho_\mathrm{s}$, and the core size is given in terms of the initial $r_\mathrm{max}$.}
    \label{fig:CoreDensityVSCoreSize}
\end{figure*}

Figure~\ref{fig:CoreDensityVSCoreSize} shows for all DM subhalos the central density vs.\ the core size. On the left side, the CDM and isotropic SIDM subhalos are shown, and on the right side, the forward-dominated SIDM subhalos are shown. The CDM subhalos (solid lines) remain roughly constant around the initial value. All the SIDM subhalos evolve very similarly during the core expansion phase; the core size increases while the central density decreases. Only the isotropic SIDM subhalo on the inner orbit (dashed green line on the left side) shows a significantly stronger decrease in the central density. As the core collapses, the SIDM subhalos show a strong increase in the central density, reaching higher values for the SIDM subhalos on the outer orbits. On the right side, the SIDM subhalos, which are destroyed by tidal forces, show an overall decreasing central density and increasing core size.

\begin{figure*}
    \centering
    \includegraphics[width=\columnwidth]{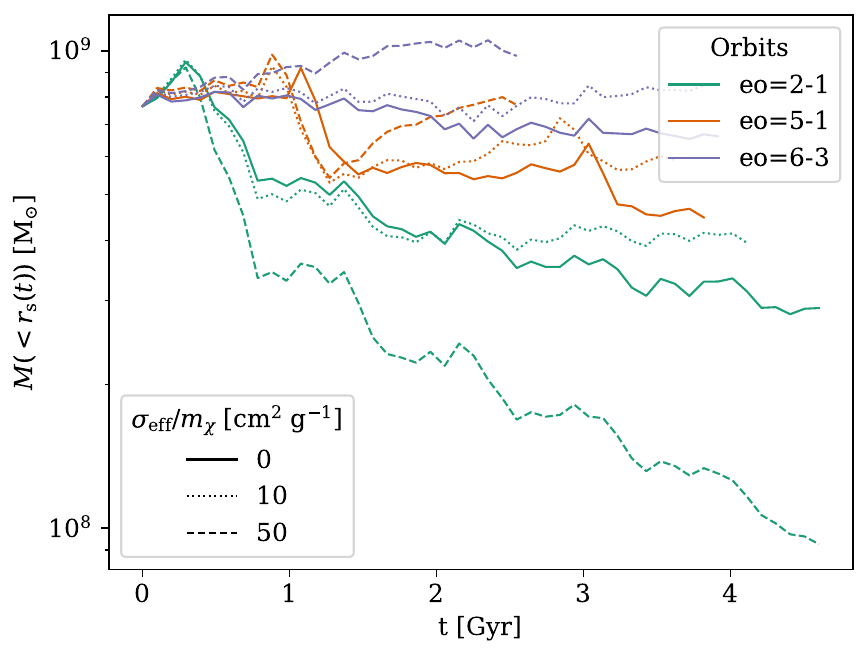}
    \includegraphics[width=\columnwidth]{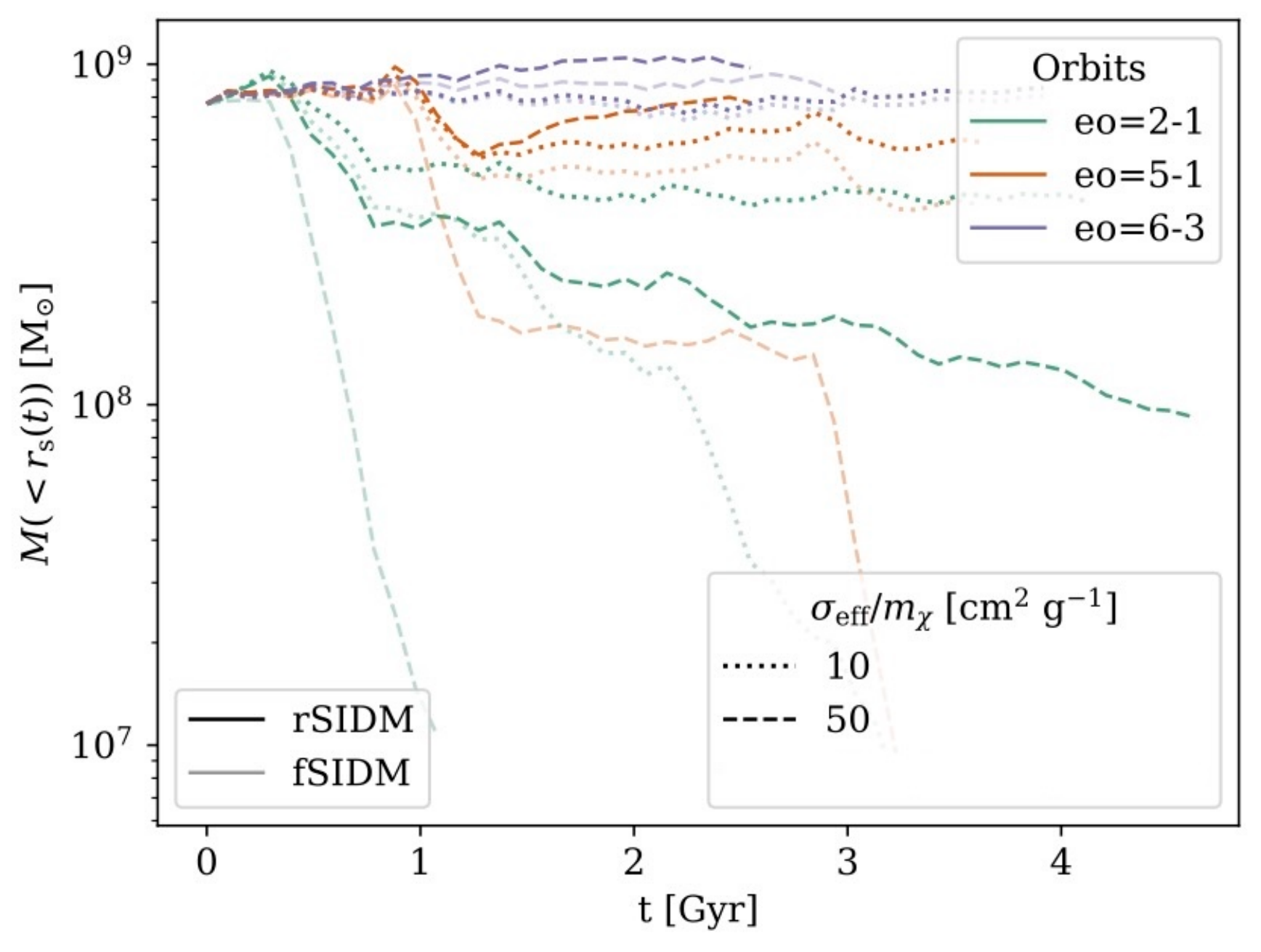}
    \caption{Enclosed mass within the scale radius. Left panel: CDM and isotropic SIDM subhalos, right panel: forward-dominated SIDM subhalos. The scale radius is calculated at each time step of the corresponding CDM subhalo (see Eq.~\eqref{eq:rho_s}).}
    \label{fig:EnclosedMass}
\end{figure*}

Figure~\ref{fig:EnclosedMass} shows the enclosed mass within the scale radius for all DM subhalos. The scale radius is calculated at each time step with Eq.~\eqref{eq:rho_s}) from the corresponding CDM subhalos. On the left, the CDM and isotropic SIDM subhalos are shown, and on the right, the forward-dominated SIDM subhalos. On the outer orbit (purple), the CDM subhalo shows a small decline in the enclosed mass, while the SIDM subhalos show an increase during core collapse. On the intermediate orbit (orange), the CDM subhalo shows an increase during pericentre passage, which is followed by a sharp decline in the enclosed mass as a result of tidal stripping. The corresponding SIDM subhalos also show an increase during the pericentre passage, followed by a sharp decline, but slightly earlier than the CDM subhalo. On the inner orbit (green), the subhalos directly show an increase in the enclosed mass. As the subhalos lose more mass on closer orbits, this results in lower enclosed masses for closer orbits. The forward-dominated SIDM subhalos have smaller enclosed masses than the corresponding isotropic SIDM subhalo.

\section{From isolation to subhalo at $z=2$}
\label{sec:z=2}
As discussed in Sect.~\ref{sec:SubhaloEvoOverview}, the DM subhalo is isolated before becoming bound to the host system. So far, we have let the DM subhalo evolve directly in the host environment. Here, we investigate the differences between our idealised simulations and a DM subhalo, which is first isolated and then placed on the inner orbit (eo=2--1).

For this, we produce the initial conditions of a similar DM subhalo, but assuming the mass-concentration relation \citep{Dutton_2014} for $z=2$.
This DM halo has a virial mass of $M_\mathrm{vir}=10^{10} \, \mathrm{M_\odot}$ and a typical concentration of $c_\mathrm{vir}=19$, corresponding to an NFW profile with $\rho_\mathrm{s}=6.4\times10^7 \ \mathrm{M}_\odot \ \mathrm{kpc}^{-3}$ and $r_\mathrm{s}=1.2 \ \mathrm{kpc}$. The DM halo is evolved in isolation for the age of the universe at $z=2$ and then placed on the inner orbit. For this, we use the smaller cross section [$\sigma_0/m_\chi = 18 \ \mathrm{cm}^2 \ \mathrm{g}^{-1}$, $w = 188 \ \mathrm{km} \ \mathrm{s}^{-1}$] with isotropic scattering. 

\begin{figure}
    \centering
    \includegraphics[width=\columnwidth]{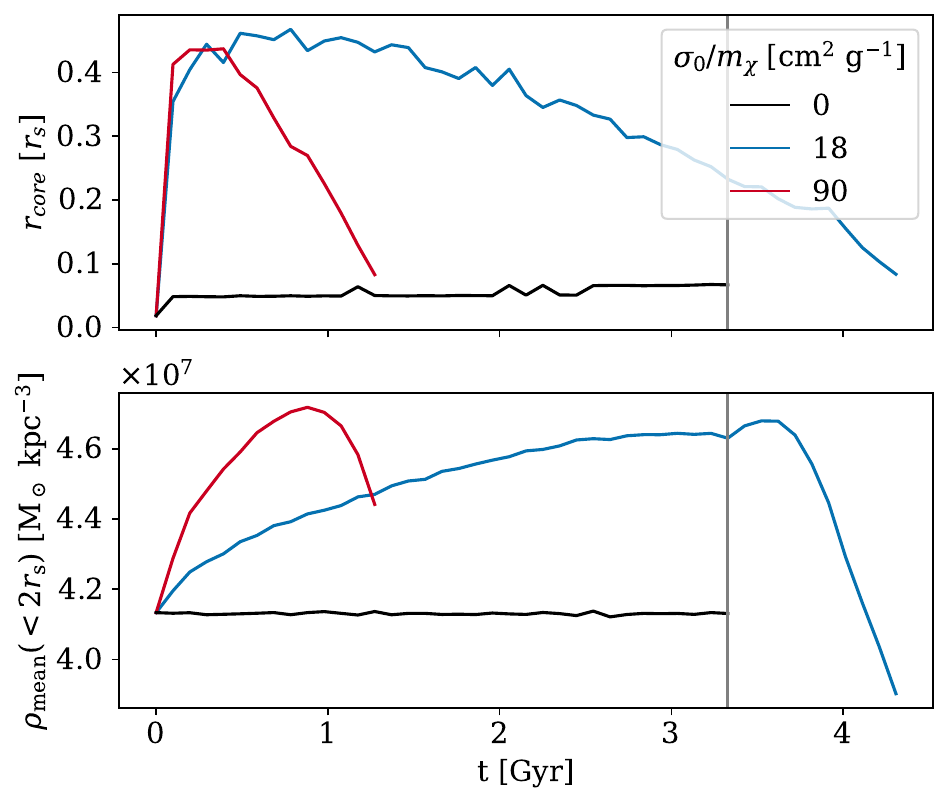}
    \caption{Evolution of the core size and mean central density within two times the initial scale radius for the rSIDM subhalos at $z=2$. The grey vertical line indicates when the isolated halo becomes a subhalo on the inner orbit (eo=2--1). The radius $r_\mathrm{core}$ is given in units of the scale radius of the initial NFW subhalo.}
    \label{fig:Core_MeanCentralDensity_fSIDM_z=2}
\end{figure}

In Fig.~\ref{fig:Core_MeanCentralDensity_fSIDM_z=2}, the evolution of the core size and the mean central density are shown. The red line is the isolated halo with the higher cross section, and the blue line is the halo with the lower cross section, which is then placed on the inner orbit. The grey vertical line indicates when the DM subhalo is placed on the inner orbit in the host system. The core size evolution seems to be slightly elongated after the grey vertical line. The mean central density suddenly increases as the DM subhalo is in the host environment.

\begin{figure}
    \centering
    \includegraphics[width=0.9\columnwidth]{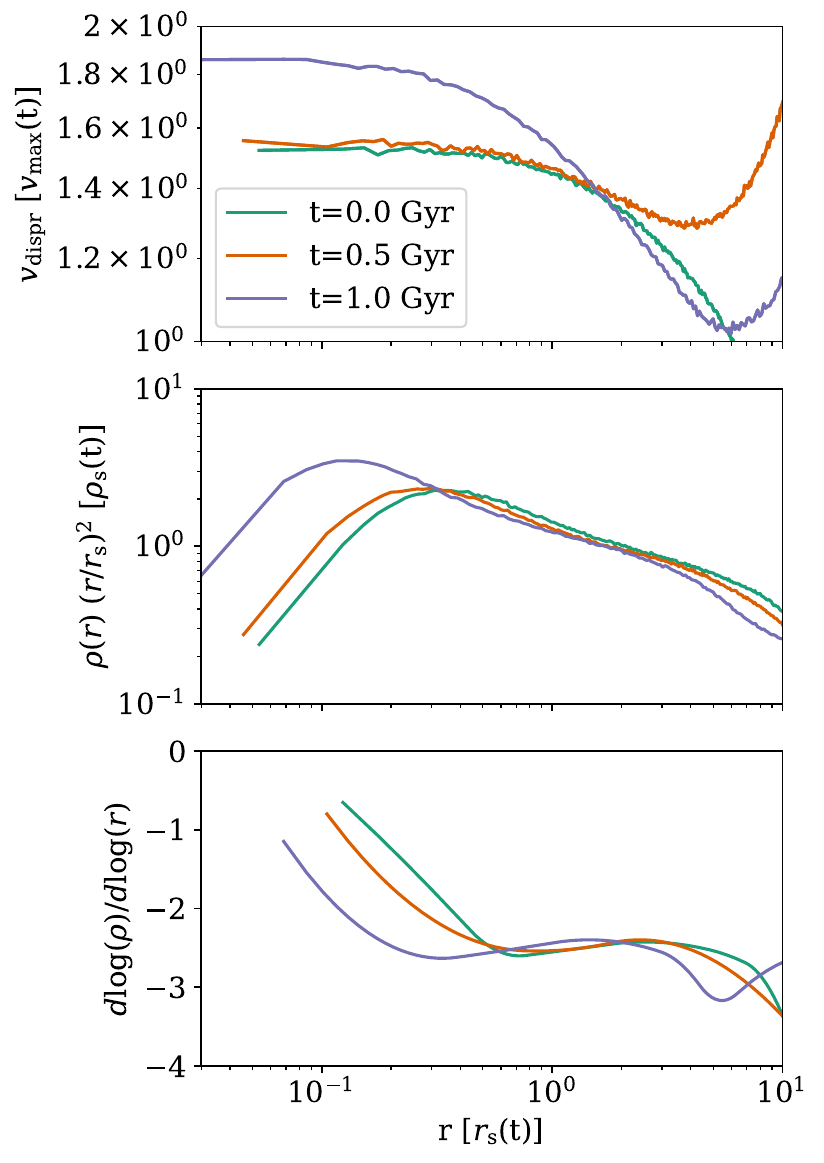}
    \caption{Evolution of the rSIDM subhalos at $z=2$ on the inner orbit (eo=2--1). The left panel shows the velocity dispersion as a function of radius. The middle panel displays the density profile of the subhalo. The right panel presents the logarithmic slope of the density profile with respect to the radius. The radius $r$ is given in units of the scale radius, which is calculated at each time step of the corresponding CDM subhalo (seen in Fig.~\ref{fig:TimeEvolution}) as $v_\mathrm{max}$ and $\rho_\mathrm{s}$.}
    \label{fig:DensityVelDisprProfileEvolution_fSIDM_z=2}
\end{figure}

In Fig.~\ref{fig:DensityVelDisprProfileEvolution_fSIDM_z=2}, the density, velocity dispersion, and density slope profiles are shown. The green line with the label $t = 0 \ \mathrm{Gyr}$ represents the profile of the DM subhalo when it is placed in the host environment. The next time step is during the pericentre, and the last time step is during the apocentre. During the pericentre, the velocity dispersion profile is overall slightly increased. The increase in the velocity dispersion profile arises from the tidal forces, which accelerate all subhalo particles. During the apocentre, the outer density and velocity dispersion profile are reduced due to the stripped particles. We can compare this subhalo to the subhalo in the right column of Fig.~\ref{fig:TimeEvolution}. The subhalo, which evolved first in isolation, shows an increase in the central density and velocity dispersion during its orbit, while the other show an decrease even after core formation. This highlights how crucial the pre-infall history and orbital path of a SIDM subhalo are for determining its subsequent evolution.

\section{Verification tests}
\label{sec:verification_tests}
This section presents a series of tests to verify the implementation of the SSHI process mechanism. We assess the individual components of the method separately. The first two tests evaluate the SSHI process process in the presence of background particles at rest. The third test examines the Eddington inversion, which is crucial for accurately reproducing the local host environment.
\subsection{Deceleration Problem}
\begin{figure}
    \centering
    \includegraphics[width=\columnwidth]{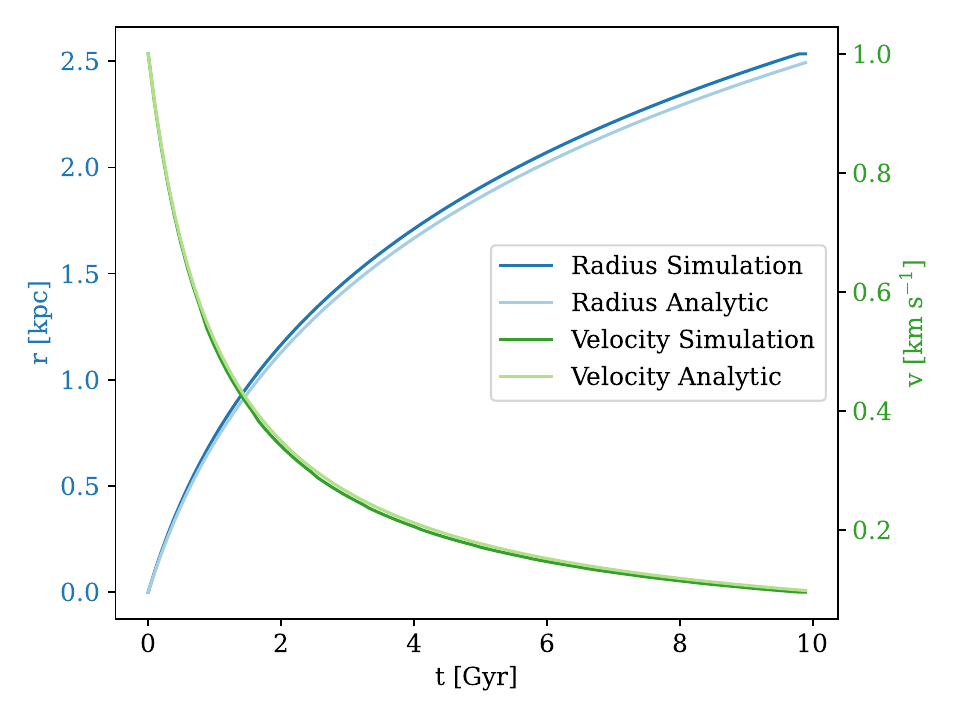}
    \caption{Comparison between the analytic solution (Eq.~\eqref{eq:AnalyticDeceleration}) and the simulation for the deceleration problem. The distance and velocity of a test particle are shown as it moves through a constant background density.
    }
    \label{fig:Deceleration-Test}
\end{figure}
The first test for the implementation of the SSHI process effect is the deceleration problem, similar to the deceleration problem in \citet{Fischer_2021a}. In this test, a single test particle moves through a background with a constant density $\rho$. In our deceleration problem, the background particles are described analytically, just as we model a host halo. The background density is sampled on the fly using virtual particles. These virtual particles are at rest.

In this scenario, we use the fSIDM module to simulate the scattering. In fSIDM, the scattering cross sections are such that the typical scattering angles are small, leading to an effective drag force. As a result, particles experience deceleration due to interactions with the surrounding matter. For this test, the perpendicular momentum diffusion is turned off; therefore, the direction of movement does not change.

We choose this scenario with fSIDM because it allows us to test our implementation against a simple analytic expression effectively. 
The analytic trajectory of the test particle, subject to the drag force, can be derived from the following equation \citep{Fischer_2021a},
\begin{equation}
    \Ddot{\vec{x}}=-\frac{1}{2}\dot{\vec{x}}^2\rho\frac{\sigma_\mathrm{\tilde{T}}}{m_\chi}
    \label{eq:AnalyticDeceleration} \,.
\end{equation}
We performed the test simulation using a background density of $\rho=4.46 \times 10^7 \ \mathrm{M_\odot} \ \mathrm{kpc}^{-3}$ and a modified momentum transfer cross section (Eq.~\eqref{eq:modified_momentum_transfer_cross_section}) of $\sigma_\mathrm{\tilde{T}}/ m_\chi=200 \ \mathrm{cm}^2\ \mathrm{g}^{-1}$. For each time step, $N_\mathrm{virt}=64$ particles of the background are sampled at the coordinates of the test particle, and then the interactions are calculated. Figure~\ref{fig:Deceleration-Test} shows the analytic and simulated solutions. The results match closely, confirming that our SSHI process implementation aligns well with the expected results.

\subsection{Wind tunnel test}
\begin{figure}
    \centering
    \includegraphics[width=\columnwidth]{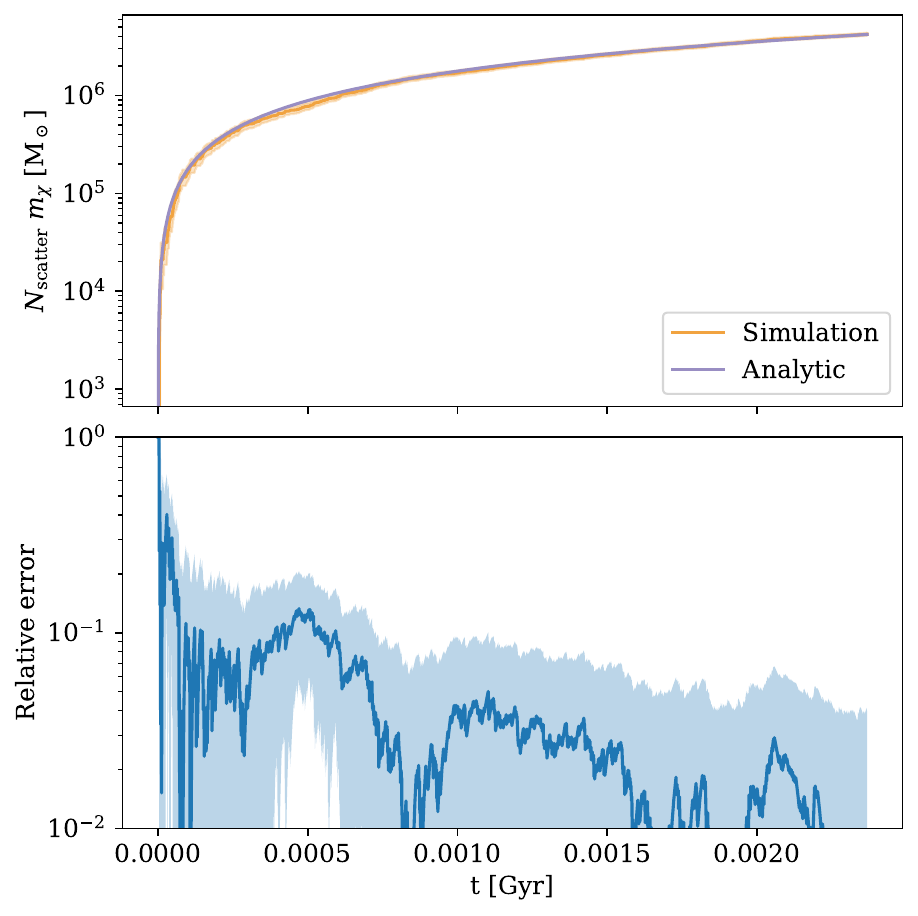}
    \caption{Top panel shows the number of scatter events $N_\mathrm{scatter}$ caused by the SSHI process between an SIDM halo and a constant DM background density $\rho$, compared between the simulation results (orange) and the analytic prediction from Eq.~\eqref{eq:NumberOfScatterEvents} (purple). The bottom panel shows the relative error between the simulation and the analytic prediction with an error band from the Poisson noise of the simulation.}
    \label{fig:ExpectedScatter}
\end{figure}
As a second test for the SSHI process implementation, we simulate a DM halo moving with a velocity $v$ through a constant background density $\rho$. The background particles are at rest and are represented by the virtual particles of our SSHI process implementation. For each DM subhalo particle, these virtual particles are sampled at every time step and represent the local background density. For this test, the rSIDM module is used with an isotropic cross section to simulate the scattering. 

This setup has an analytic estimate for the scattering rate per halo particle,
\begin{equation}
    R = \frac{\sigma}{m_\chi} \, v \, \rho \,.
\end{equation}
The scattering rate depends on the density of the background $\rho$, the velocity of the halo $v$, and the cross section $\sigma/m_\chi$. The actual velocity for the scattering between particles deviates from the velocity of the halo $v$ because of the velocity dispersion inside the halo. The error arising from this approximation vanishes in the limit where the velocity $v$ of the halo relative to the background becomes large compared to the velocity dispersion of the halo. 

By multiplying the scattering rate $R$ with the total number of halo particles $N$, and a time interval ${\Delta t}$, we get the expected number of scatter events $N_\mathrm{scatter, \ num}$ in a given time step,
\begin{equation}
    N_\mathrm{scatter, \ num} = N \, R \, {\Delta t} \,.
    \label{eq:NumberOfScatterEvents}
\end{equation}
By multiplying it by the numerical particle mass $m$, we get the mass of the subhalo that scatters with the background,
\begin{equation}
    N_\mathrm{scatter} \ m_\chi = N_\mathrm{scatter, \ num} \ m \,,
    \label{eq:ScatteredMass}
\end{equation}
where $m_\chi$ is the physical mass of one DM particle.
We performed the test simulation using a background density of $\rho = 2.46 \times 10^4 \ \mathrm{M_\odot} \ \mathrm{kpc}^{-3}$. The $N$ particles are sampled with \textsc{SpherIC} \citep{Garrison-Kimmel:2013yys, 10.1111/j.1365-2966.2008.13126.x} for $N = 2 \times 10^6$ particles, $r_\mathrm{s}=2.6 \ \mathrm{kpc}$ and $\rho_\mathrm{s}=2.4\times10^7 \ \mathrm{M_\odot} \ \mathrm{kpc}^{-3}$. The self-interaction cross section is $\sigma/m_\chi = 50 \ \mathrm{cm}^2 \ \mathrm{g}^{-1}$ and the halo has an initial velocity of $v = 686 \ \mathrm{km} \ \mathrm{s}^{-1}$. For each time step, $N_\mathrm{virt}=48$ particles of the background are sampled at the coordinates of each DM halo particle, and then the potential scatter is calculated. In Fig.~\ref{fig:ExpectedScatter}, the analytic and simulated scattered masses are shown and their relative error with an error band from the Poisson noise of the simulation. The simulation closely follows the analytic solution with almost no significant deviation. This confirms that our SSHI process implementation aligns well with the expected results.

\subsection{Eddington Inversion}
\label{sec:VeriEddingtonInversion}
Here, we present a two-step validation of the Eddington inversion method as implemented in our simulation framework \citep{1916MNRAS..76..572E}. The goal is to ensure that the sampling procedure accurately reflects the underlying distribution function derived from a known density profile.
First, we test the sampling routine by reproducing the velocity dispersion profile for a halo following an NFW profile. The sampled profile is compared to the analytic result to assess accuracy across a broad radial range.
Second, we embed the sampling routine within a scattering simulation of a DM subhalo, verifying the method’s accuracy in a dynamic setting. The host system is described analytically by an NFW profile. The velocities of the virtual host particles are sampled during the scattering process and used to compare the produced velocity distribution to the analytic Maxwell-Boltzmann distribution.

\begin{figure}
    \centering
    \includegraphics[width=\columnwidth]{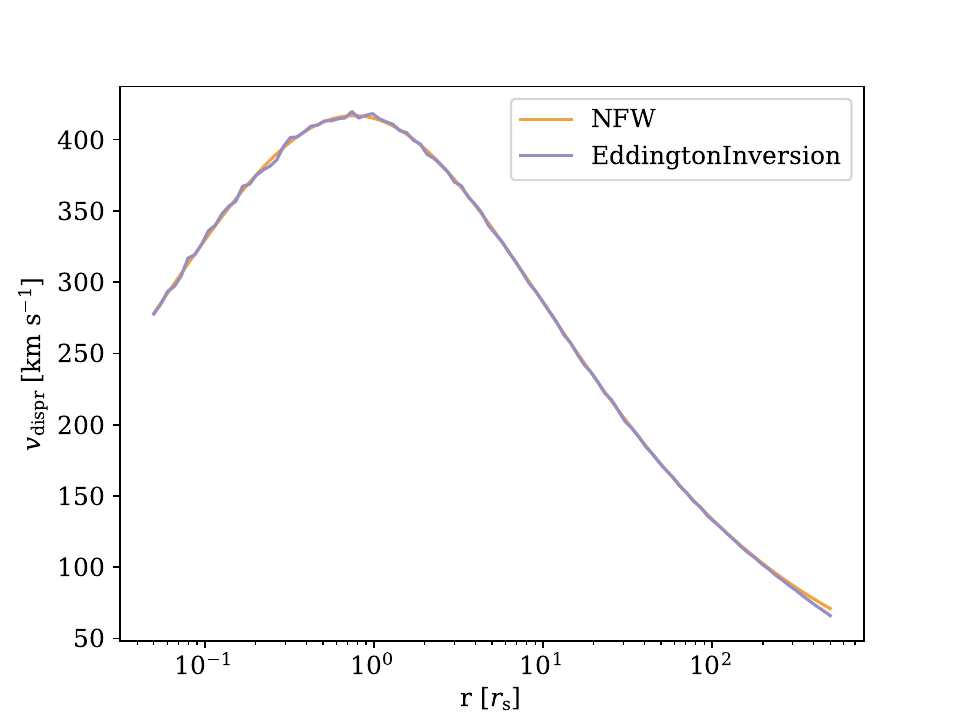}
    \includegraphics[width=\columnwidth]{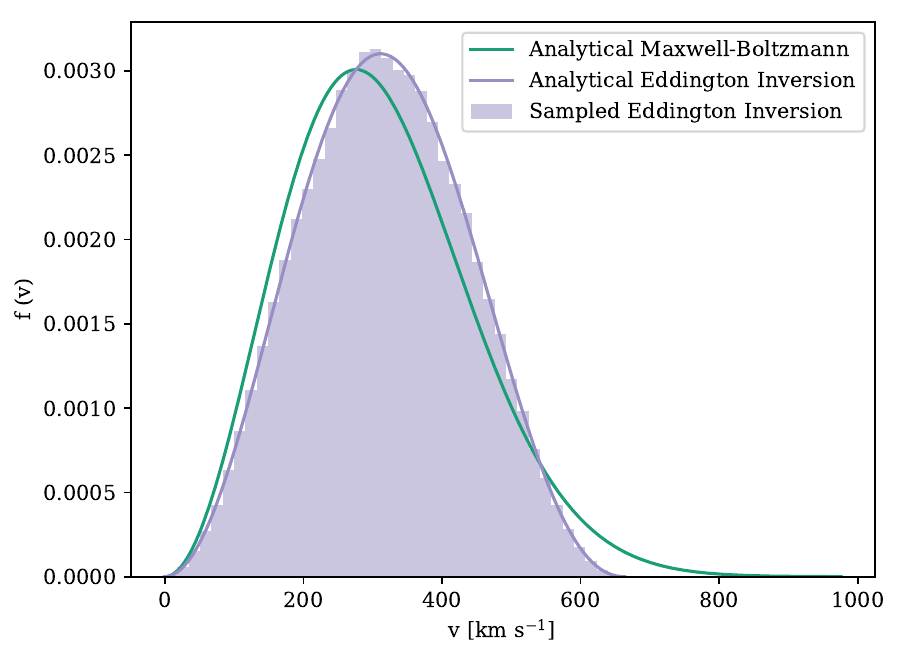}
    \caption{Top panel: Comparison between the produced velocity dispersion via the Eddington Inversion (purple) and the analytic NFW velocity dispersion (orange). Bottom panel: Velocity distribution of sampled particles at $5 \ r_\mathrm{s}$ via Eddington Inversion (purple) compared to Maxwell-Boltzmann distribution (green).}
    \label{fig:EddingtonInversion_0}
\end{figure}
The accuracy of our implemented sampling method is first tested by reproducing a complete velocity dispersion profile. This velocity dispersion profile follows an NFW profile with the parameters: $r_\mathrm{s}=50.67 \ \mathrm{kpc}$ and $\rho_\mathrm{s}= 1.11 \times10^6 \ \mathrm{M}_\odot \ \mathrm{kpc}^{-3}$. 
For this test, particles are sampled within the radial range $0.05 \ r_\mathrm{s}$ to $500 \ r_\mathrm{s}$, distributed across $100$ logarithmically spaced radii. In each bin, $10^4$ particles are sampled, and their velocity dispersion is calculated.
In Fig.~\ref{fig:EddingtonInversion_0}, the sampled velocity dispersion profile is compared to the analytic profile. We observe that the sampling routine reproduces the velocity dispersion profile very well. Significant differences appear only at large radii.

Next, we test the sampling method within the scattering routine. We start a simulation of a DM subhalo with the SSHI process implemented. This DM subhalo orbits circularly at a radius of $5 \ r_\mathrm{s}$, where $r_\mathrm{s}$ is the host halo’s scale radius. The host halo has the same properties as for the first test of the Eddington Inversion.
We collect the velocities of the virtual host particles that are sampled as part of the scattering routine. 
In Fig.~\ref{fig:EddingtonInversion_0}, the velocity distribution of these virtual host particles is compared to the analytic Maxwell-Boltzmann distribution,
\begin{equation}
    f(v)=\sqrt{\frac{2}{\uppi}}\frac{v^2}{\vdisp^3}\exp{\left(-\frac{v^2}{2\vdisp^2}\right)} \,.
\end{equation}
The latter is shown using the same one-dimensional velocity dispersion $\vdisp$, as that measured from the sampled velocity dispersion. 
For this test, more than $10^6$ virtual particles were sampled. The figure shows that the Maxwell-Boltzmann distribution includes a high-velocity tail, which is absent in the Eddington inversion. This tail means there are more fast-moving particles, which makes subhalos sampled from the Maxwell-Boltzmann distribution less stable. These high-velocity particles are not gravitationally bound and are gradually ejected from the CDM subhalo. Therefore, modelling the SSHI process is more accurate when virtual particles are sampled using the Eddington inversion.

\end{appendix}

\end{document}